\documentclass[aps, prb, reprint,twocolumn, showpacs, superscriptaddress, amsfonts, amsmath, amssymb,floatfix,longbibliography]{revtex4-1}
\usepackage{dcolumn}
\usepackage{multirow}
\usepackage{amsmath}
\usepackage{graphicx,epsfig,psfrag}
\usepackage{amssymb}
\usepackage{natbib}
\usepackage[colorlinks,linkcolor={magenta},citecolor={magenta},urlcolor={magenta},breaklinks=true]{hyperref}
\usepackage{bookmark}

\usepackage[utf8]{inputenc}
\usepackage[T1]{fontenc}
\usepackage{lmodern}

\usepackage{graphicx}
\graphicspath{{./images/}}

\usepackage{color}
\definecolor{darkgreen}{rgb}{0,0.5,0}
\definecolor{darkblue}{rgb}{0,0,0.5}
\definecolor{purple}{rgb}{0.35,0,0.35}
\definecolor{orange}{rgb}{1,0.5,0}
\definecolor{grey}{rgb}{.6,.6,.6}

\definecolor{forestgreen}{rgb}{0.13, 0.55, 0.13}

\newcommand{\change}[1]{{\textcolor{black}{#1}}}

\usepackage{bbm}

\usepackage{hyperref}

\newcommand{\rem}[1]{}

\newcommand{\veps}{\varepsilon}

\DeclareMathOperator{\sech}{sech}

\begin{document}

\title{The Ohmic two-state system from the perspective of the interacting resonant level model: Thermodynamics and transient dynamics}

\author{H. T. M. Nghiem}
\affiliation{Peter Grünberg Institut and Institute for Advanced Simulation, Research Centre Jülich, D-52425 Jülich, Germany}
\author{D. M. Kennes}
\affiliation{Institut für Theorie der Statistischen Physik, RWTH Aachen University and 
JARA—Fundamentals of Future Information Technology, 52056 Aachen, Germany}
\author{C. Kl\"ockner}
\affiliation{Institut für Theorie der Statistischen Physik, RWTH Aachen University and 
JARA—Fundamentals of Future Information Technology, 52056 Aachen, Germany}
\author{V. Meden}
\affiliation{Institut für Theorie der Statistischen Physik, RWTH Aachen University and 
JARA—Fundamentals of Future Information Technology, 52056 Aachen, Germany}
\affiliation{Peter Grünberg Institut, Research Centre Jülich, D-52425 Jülich, Germany}
\author{T. A. Costi}
\affiliation{Peter Grünberg Institut and Institute for Advanced Simulation, Research Centre Jülich, D-52425 Jülich, Germany}

\date{\small\today}

\begin{abstract}
We investigate the thermodynamics and transient dynamics
of the (unbiased) Ohmic two-state system by exploiting the equivalence of this
model to the interacting resonant level model. 
For the thermodynamics, we show, by using the numerical renormalization group
(NRG) method, how the universal specific heat and susceptibility
curves evolve with increasing dissipation strength, $\alpha$, from those of an isolated two-level 
system at vanishingly small dissipation strength, with the characteristic
activated-like behavior in this limit, to those of the isotropic 
Kondo model in the limit $\alpha\to 1^{-}$. At any finite $\alpha>0$, and for
sufficiently low temperature, the behavior of the thermodynamics is
that of a  gapless renormalized Fermi liquid. Our results compare well with available Bethe ansatz 
calculations at  rational values of $\alpha$, but go beyond these,
since our NRG calculations, via the interacting resonant level model,
can be carried out efficiently and accurately for {\em arbitrary} 
dissipation strengths $0\leq \alpha< 1^{-}$. We verify the dramatic
renormalization of the low-energy thermodynamic scale $T_{0}$ with
increasing $\alpha$, finding excellent agreement between NRG
and density matrix renormalization group (DMRG) approaches.
For the zero-temperature transient  dynamics of the two-level system, $P(t)=\langle \sigma_{z}(t)\rangle$,
with initial-state preparation $P(t\leq0)=+1$, we apply the
time-dependent extension of the  NRG (TDNRG) to the interacting
resonant level model, and compare the results obtained with those from
the noninteracting-blip approximation (NIBA), the functional
renormalization group (FRG), and the time-dependent density 
matrix renormalization group (TD-DMRG). We demonstrate 
excellent agreement on short to intermediate time scales between TDNRG 
and TD-DMRG for $0\lesssim\alpha \lesssim 0.9$ for $P(t)$, and between TDNRG 
and FRG in the vicinity of $\alpha=1/2$. Furthermore, we quantify the error in the 
NIBA for a range of $\alpha$, finding significant errors in the latter even for $0.1\leq \alpha\leq
0.4$. We also briefly discuss why the long-time errors in the present
formulation of the TDNRG  prevent an investigation of the crossover
between coherent and incoherent dynamics. Our results for $P(t)$ at
short to intermediate times could act as useful benchmarks for the development 
of new techniques to simulate the transient dynamics of spin-boson problems.
\end{abstract}


\maketitle

\section{introduction}
\label{sec:intro}
The Ohmic two-state model describes a quantum mechanical system tunneling between two states and subject to a  
coupling to environmental degrees of freedom \cite{Caldeira1983,Leggett1987}. It is a ubiquitous model
in condensed matter physics, capable of
describing, to varying degrees of approximation, the low-energy physics of a wide range of systems, including, 
for example, the tunneling of defects in solids \cite{Zimmerman1991,Golding1992}, 
the diffusion of protons and muons in metals
\cite{Kondo1984,Wipf1984,Grabert1997}, or the quantum mechanical
tunneling of fluxoid states in superconducting quantum interference
devices\cite{Leggett1984}. Other possible realizations that have been
proposed include two-level atoms in optical fibers \cite{LeClair1997}
and, more recently, Cooper pair boxes, acting as two-level systems, and coupled to an
electromagnetic environment consisting of an array of Josephson
junctions\cite{LeHur2012,Goldstein2013,Peropadre2013,Snyman2015,Bera2016,Gheeraert2016}. 
Cooper pair boxes can host qubits with long coherence times
$T_{2}\sim 10-20\mu s$, and are currently being \change{investigated\cite{Paik2011}} as one
possible alternative to solid-state qubits\cite{Loss1998} in the context of
quantum information processing. The model also provides a microscopic starting
point for describing electron transfer between donor and acceptor
molecules in photosynthesis and other biological processes 
\cite{Marcus1985,Tornow2008}. For an overview of the Ohmic two-state
system, and quantum dissipative systems in general, see
Ref.~\onlinecite{Weiss2008}.

The Hamiltonian of the Ohmic two-state system, or the Ohmic spin-boson
model (SBM), terms which we shall use interchangeably, is given by,
\begin{align}
& H_{\rm SBM} =
\underbrace{-\frac{1}{2}\Delta_{0}\sigma_{x}+\frac{1}{2}\epsilon\sigma_{z}}_{H_{\rm 
  TLS}} +
\underbrace{\frac{1}{2}\sigma_{z}\sum_{i}\lambda_{i}(a_{i}+a_{i}^{\dagger})}_{H_{\rm 
  int}}\nonumber\\ 
& +
\underbrace{\sum_{i}\omega_{i}(a^{\dagger}_{i}a_{i}+1/2)}_{H_{\rm bath}}.\label{eq:sbm}
\end{align}
``Ohmic'' refers to a particular choice of the couplings $\lambda_i$ and oscillator frequencies $\omega_i$, 
that we shall specify below. The first term $H_{\rm TLS}$ describes
a two-level system with bias splitting $\epsilon$ and bare tunneling amplitude $\Delta_{0}$, and
$\sigma_{i=x,y,z}$ are Pauli spin matrices. The third term, $H_{\rm
  bath}$, is the environment and consists of
an infinite set of harmonic oscillators ($i=1,2,\dots,\infty$) with $a_{i} (a_{i}^{\dagger})$ the
annihilation (creation) operators for a harmonic oscillator of frequency $\omega_{i}$ and 
$0\le \omega_{i}\le \omega_{\rm c}$, where $\omega_{\rm c}$ is an upper cutoff frequency. The noninteracting
density of states of the environment is denoted by $g(\omega)=\sum_{i}\delta(\omega-\omega_{i})$ and
is finite in the interval $[0,\omega_{\rm c}]$ and zero otherwise.
Finally, $H_{\rm int}=\frac{1}{2}\sigma_{z}\sum_{i}\lambda_{i}(a_{i}+a_{i}^{\dagger})$ describes
the coupling of the two-state system co-ordinate $\sigma_{z}$ to the oscillators, with
$\lambda_{i}$ denoting the coupling strength to oscillator $i$. The function 
$\Gamma(\omega+i\delta)=\sum_{i}\lambda_{i}^2/(\omega-\omega_{i}+i\delta)=
\int d\omega' \lambda(\omega')^2\,g(\omega')/(\omega-\omega'+i\delta)$ 
characterizes the system-environment interaction. In particular, the  spectral function
$J(\omega)=-\frac{1}{\pi}{\rm Im} [\Gamma(\omega)]=\sum_{i}\lambda_{i}^2\delta(\omega-\omega_i)$ 
allows a classification of models into sub-Ohmic, Ohmic and super-Ohmic depending on whether
the low-frequency behavior of $J(\omega\to 0)\sim \omega^s$ is sublinear ($s<1$), linear
($s=1$), or superlinear ($s>1$).

In this paper,  we  shall be interested in the case of Ohmic dissipation ($s=1$), characterized
by a spectral function $J(\omega)$ linear in frequency, $J(\omega)=
2\pi\alpha \omega\theta(\omega_c - \omega)$, with $\alpha$ the dimensionless dissipation strength
describing the strength of the coupling of the TLS to its
environment. The  Ohmic case is particularly interesting because,
aside from its relevance to a large number of physical
situations \cite{Weiss2008}, \change{it also allows for a mapping},
in the scaling limit $\Delta_{0}/\omega_c\ll 1$, to a number
of interesting fermionic models, including the anisotropic Kondo model (AKM), the
spin-fermion model (SFM), and the interacting resonant level model (IRLM)
\cite{Guinea1985,Vigman1978,Costi1999}. Such equivalences prove to be 
very useful because powerful techniques developed for the fermionic models,
such as the numerical renormalization group (NRG), can be used
to investigate diverse properties of the Ohmic SBM, without restriction
on temperature, coupling strength, or other system parameters. 

More specifically, we shall use the equivalence of the Ohmic two-state system to the IRLM in
order to, (a), show explicitly that the IRLM can recover all the
interesting regimes of the Ohmic two-state system, from zero
($\alpha=0$) to maximum dissipation strength ($\alpha\to
1^{-}$),\footnote{We do not address the regime $\alpha>1$, since in
  this regime, quantum mechanical tunneling is absent for most of the
  parameter space, a situation that can be addressed with perturbative
methods.},
(b), shed further light on the thermodynamic properties of the former by application of the NRG to
the latter, and, (c), shed light on the accuracy of the recently developed time-dependent extension
of the numerical renormalization group (TDNRG) for transient
quantities\cite{Anders2005,Anders2006,Nghiem2014a,Nghiem2014b} by
comparison with complementary approaches, such as the TD-DMRG, the
\change{functional renormalization group} (FRG), and the noninteracting-blip approximation (NIBA). For our
purposes, a study of the unbiased Ohmic two-state system
[i.e., $\epsilon=0$ in Eq.~(\ref{eq:sbm})] suffices. Apart from the NIBA, which yields unphysical
results for transient dynamics at finite bias \cite{Weiss2008}, the
other methods can equally well be applied to finite bias also.

Historically, the main interest in the Ohmic two-state system has been
to understand the crossover from coherent to incoherent dynamics
of the two-level system as the coupling to the environment is
increased \cite{Caldeira1983,Leggett1987,Weiss2008,Kennes2013b}. Less attention has been given to thermodynamic properties,
(see Refs.~\onlinecite{Goerlich1988,Kato2000,Costi1998,Costi1999}).
While previous calculations for specific heats and dielectric susceptibilities
of the Ohmic two-state system are available via the NRG and Bethe
ansatz applied to the AKM\cite{Costi1998,Costi1999}, the results obtained using these methods
remain incomplete: previous NRG calculations used a coarse
temperature grid, such that the activated behavior of the specific heat for $\alpha\to 0$ was not well captured\cite{Costi1998}, while the
Bethe ansatz results were readily available only at certain rational
values of  the dissipation strength $\alpha=1/\nu$ and
$\alpha=1-1/\nu$ for $\nu=2,3,4,\dots$ due to the complexity of the
thermodynamic Bethe ansatz
equations for general $\alpha$ \cite{Costi1999, Tsvelick1983b}. 
By implementing recent advances in the NRG \cite{Campo2005,Merker2012b}, we are
able to use a much finer temperature grid, thereby resolving all
features in the specific heat with sufficient accuracy, even in the limit $\alpha\to 0$ (see 
Sec.~\ref{subsubsec:specific-heats}).
More importantly, the equivalence of the IRLM to the Ohmic two-state system (\ref{eq:sbm}) allows 
calculations at {\em arbitrary} $\alpha$ to be carried out straightforwardly, in contrast to
the limited values of $\alpha$ that are easily achievable within the Bethe ansatz. 
Thus, the IRLM results presented in this paper fill a gap in our quantitative
understanding of the thermodynamics of the Ohmic two-state system. 
In addition, they demonstrate that the equivalence between the IRLM
and the Ohmic two-state system is indeed valid for the whole range of
dissipation strengths of interest. That such calculations are
possible and yield meaningful results in the limit $\alpha\to 0$ is  not immediately obvious, since this
limit corresponds to infinite Coulomb interaction in the IRLM (see
Sec.~\ref{sec:model} and Appendix~\ref{appendix:equivalences}).
Indeed, we shall show that the IRLM can describe the evolution of
the thermodynamic properties of the Ohmic two-state system from the limit of an isolated
two-level system at $\alpha=0$ to the limit  of strong dissipation 
$\alpha\rightarrow 1^{-}$ where the universal scaling functions become
those of the (isotropic) Kondo model. At the technical level, a further advantage
in carrying out NRG calculations on the IRLM as opposed to the AKM (or
the SFM) is that the former is spinless. This implies that a larger number of states can 
be retained within an NRG treatment of the IRLM than for the AKM (or
the SFM),  thereby allowing highly accurate results to
be obtained for all dissipation strengths; the accuracy of these results will be 
demonstrated by comparison with limiting cases and Bethe ansatz
calculations at representative values of $\alpha$. 

Beyond presenting  results for thermodynamics at general values of $\alpha$, and
demonstrating that the IRLM can be applied to describe all regimes of
interest of the Ohmic two-state system, the other
main aim of  thist study is to shed light on the accuracy 
of the TDNRG for zero-temperature transient quantities at ``short'' to ``intermediate
times'' (terms that we shall define precisely below). We do not
address in any detail the $t\to 0^{+}$ limit of transient quantities, which is known
to be exact in the TDNRG \cite{Nghiem2014a}, nor the approach 
to the long-time limit $t\to\infty$ of transient quantities where
finite errors appear (described in detail elsewhere\cite{Nghiem2014a,Nghiem2014b}), but focus our attention  on
the accuracy of the TDNRG in the short to intermediate time range
specified below. Such tests have so far been limited to cases with
exact solutions \cite{Anders2006,Nghiem2014a,Nghiem2014b}. It is
therefore of some interest to assess the method for generic cases, such as for
the Ohmic two-state system at  arbitrary $\alpha>0$. For the latter, we
shall focus attention on the quantity $P(t>0)=\langle\sigma_{z}(t)\rangle$,
with initial-state preparation $\sigma_z(t\leq 0)=+1$, and assess the 
accuracy of the TDNRG results on short to intermediate time scales, i.e., on 
time scales $t\ll 1/\Delta_{\rm eff}(\alpha)$ and $t\sim 1/\Delta_{\rm 
  eff}(\alpha)$, where $1/\Delta_{\rm   eff}(\alpha)$ is a time scale 
that enters the transient dynamics within the NIBA
[see Eq.~(\ref{eq:delta_eff}) and Sec.~\ref{subsubsec:comparison-niba}]. We distinguish here
  between intermediate times $t\sim 1/\Delta_{\rm eff}(\alpha)$ and
  long times $t\gg 1/\gamma_{r}(\alpha)$, where $\gamma_r(\alpha)$ sets the
  overall decay rate of $P(t)$. The decay rate $\gamma_{r}(\alpha)$ is of order
  $\alpha\Delta_{\rm eff}(\alpha)$\cite{Weiss2008}, implying that,
for most $\alpha$, except for $\alpha\ll 1$, the intermediate time scale
$1/\Delta_{\rm eff}(\alpha)$ is also the relevant time scale for the
approach to the long-time limit. For $\alpha\ll 1$, however, times $t$ such 
that \change{$1/\Delta_{\rm eff}(\alpha)\ll t <1/\gamma_{r}(\alpha)$}, should be regarded as 
intermediate times\footnote{In general, by $\Delta_{\rm 
      eff}(\alpha)$ or $\gamma_{r}(\alpha)$ we mean the NIBA
    expressions for these quantities. However, in cases where the NIBA
    expressions become quantitatively inaccurate, such as at strong dissipation
    $\alpha>1/2$, we shall point this out.}.
We shall  directly compare TDNRG results with corresponding results
from other approaches, such as with the NIBA, which is expected to be
accurate at $\alpha\ll 1$ (on the time scales given above), with the
functional renormalization group (FRG) (which is a controlled around $\alpha=1/2$),
and with the time-dependent density matrix renormalization group (TD-DMRG)
(valid for general $\alpha$). 
Beyond demonstrating convincingly that the TDNRG is quantitatively accurate in the above time-range and for
the whole range of dissipation strengths $0\lesssim \alpha\lesssim 0.9$, 
the comparisons will also allow us to quantify the errors in the
NIBA in the regime where it is traditionally expected to be a
reasonable approximation, i.e., for $0\lesssim\alpha\lesssim 1/2$ and
$\epsilon=0$.  While many techniques have been developed to simulate the
time-dependent dynamics of the spin-boson model
\cite{Egger1994,Makarov1994,Makri1995a,Makri1995b,Wang2008,Weiss2008b,Anders2006,Orth2013}, a
quantitative test of the accuracy of the NIBA at zero temperature and
in the above regime and time-range has so far been lacking.  We provide such a test by demonstrating quantitative
differences between the TDNRG and the NIBA in the range $0.1\lesssim
\alpha\lesssim 0.4$ for times comparable to $1/\Delta_{\rm
  eff}(\alpha)$, where very much smaller differences are found between
TDNRG and TD-DMRG. We shall also briefly discuss why the errors in the
 long-time limit, $t\gg 1/\gamma_{r}(\alpha)$, of the TDNRG prevent
 an investigation of the recently revealed novel scenario
for the crossover between coherent and incoherent dynamics upon
increasing $\alpha$\cite{Kennes2013b,Kennes2013c,Kashuba2013}.

We note that although the NRG has been developed \cite{Bulla2003,Bulla2005} to deal directly also
with models such as (\ref{eq:sbm}), where an impurity couples to a
continuous bath of bosons (and possibly also to a fermionic bath), and applied to  a number
of such models \cite{Bulla2005,Anders2006,Glossop2005,Glossop2007,Chung2007,Tornow2008,Orth2010b},
this ``bosonic NRG'' is not expected, in general, to be as accurate as the NRG applied to equivalent fermionic models. 
The reason for this is that each bosonic orbital in Eq.~(\ref{eq:sbm}) can accommodate an infinite number of
bosons, whereas fermionic orbitals accommodate only a single fermion (of given spin). Consequently, the truncation of the 
Hilbert space, which is inherent in the NRG procedure, is a more
severe approximation for bosonic systems than for the
equivalent fermionic ones. Indeed, we shall see in
Sec.~\ref{subsec:thermodynamics} that available bosonic NRG results for the
specific heat of the Ohmic SBM, while qualitatively correct, are in quantitative disagreement with
those obtained from both the IRLM and the Bethe ansatz. Therefore, our highly accurate
results for thermodynamic properties of the Ohmic two-state system via
the IRLM, could act as motivation for future improvements of the bosonic NRG.

Finally, aside from the interest in the IRLM for applications to the
Ohmic two-state system\cite{LeHur2008}, the model is of
interest in its own right\cite{Vigman1978,Oliveira1981,Borda2008,Kiss2013}. 
The two-lead IRLM is of interest in understanding the role of
interactions in the linear and non-linear transport through
correlated quantum dots
\cite{Borda2007,Doyon2007,Boulat2008a,Boulat2008b,Nishino2009,Karrasch2010a,Andergassen2011,Karrasch2011,Kennes2012a,Kennes2013a,Carr2015,Nishino2015},
while the IRLM with multiple channels has recently been proposed as a starting point to explain the
peculiar heavy-fermion state of some Sm skutterudite compounds \cite{Kiss2015}.

The outline of the paper is as follows. In Sec.~\ref{sec:model} we describe the IRLM and
its equivalence to the Ohmic two-state system and discuss some
limiting cases. Details of the equivalence of the Ohmic two-state
system to a number of fermionic models, including the IRLM, via
bosonization may be found in
Appendix~\ref{appendix:equivalences}. Methods are briefly described in
Sec.~\ref{sec:methods}. In Sec.~\ref{subsec:thermodynamics}, 
we present NRG results for thermodynamic properties (specific heats,
susceptibilities, and Wilson ratios), while in Sec.~\ref{subsec:transient-dynamics}, we present
our TDNRG results for the time-dependent quantity $P(t)$, comparing them with the
NIBA (in Sec.~\ref{subsubsec:comparison-niba}), the TD-DMRG
(in Sec.~\ref{subsubsec:comparison-td-dmrg}) and the FRG
(in Sec.~\ref{subsubsec:comparison-frg}), with additional supportive results in
Appendix \ref{appendix:additional-results}. We summarize with an outlook for future work in Sec.~\ref{sec:conclusions}

\section{Interacting resonant level model and connection to the Ohmic two-state system}
\label{sec:model}
The interacting resonant level model (IRLM) is given by the following
Hamiltonian

\begin{align}
&H_{\rm IRLM}=\underbrace{\veps_{d}n_{d}+
V(f^{\dagger}_{0}d+d^{\dagger}f_{0})}_{H_{\rm   imp}}
+\underbrace{U(n_{d}-1/2)(n_{0}-1/2)}_{H_{\rm   int}}\nonumber\\
&+\underbrace{\sum_{k}\epsilon_{k}c^{\dagger}_{k}c_{k}}_{H_{\rm 
  bath}}.\label{eq:IRLM}
\end{align} 
It describes a spinless resonant level with energy $\veps_{d}$
hybridizing with a spinless bath of conduction electrons (where we wrote
$f_{0}=\sum_{k}c_{k}$ and $n_{0}=f_{0}^{\dagger}f_{0}$) and interacting with the
latter via a Coulomb interaction $U$.
The correspondence of the Ohmic SBM to this model  is given by
$\Delta_{0}=2V$, $\epsilon=\veps_d$, and
$\alpha=(1+2\delta/\pi)^2/2$ where $\delta=\arctan(-\pi\rho
U/2)$ and $\rho=1/2D$ is the density of states of the
spinless conduction electrons with half-bandwidth $D$ and the
high-energy cutoffs of the two models are related by $\omega_c=2D$. 
In the scaling limit $V/D=\Delta_{0}/\omega_c \ll 1$, the equivalence between the models can be shown via
bosonization \cite{Guinea1985}, and is valid for $-\infty \le U\le +\infty$
(describing the sector $2 \ge \alpha \ge 0$) (see
Appendix~\ref{appendix:equivalences} for a detailed derivation of this and
related equivalences). In order to set some notation, we consider
the limit of an isolated two-level system. First, note that the two states in the IRLM which comprise the 
two-level system in the SBM are the states $|\uparrow\rangle =|1\rangle_{d}|0\rangle_{0}$
and $|\downarrow\rangle =|0\rangle_{d}|1\rangle_{0}$, split by  $\epsilon=\veps_{d}$, and connected by the 
hybridization $V$ which acts as the tunneling 
term in the SBM with $V=\Delta_0/2$. States $|0\rangle_{d}
|0\rangle_{0}$ and $|1\rangle_{d}|1\rangle_{0}$ lie $U/2$ higher in
energy and become decoupled, together with the band (except for the
Wannier orbital $f_{0}$), in the limit
$U\to \infty$. Hence, the isolated two-level system
($\alpha=0$) corresponds to $U=+\infty$ in the IRLM and the
eigenvalues are given by  $E_{\pm}=\pm\frac{1}{2}\sqrt{\epsilon^2+\Delta_0^2}$.
The corresponding eigenstates, for $\epsilon=0$, are given by
$\Psi_{\pm}=\frac{1}{\sqrt{2}} (|\uparrow\rangle \pm |\downarrow\rangle)$.
From the partition function
$Z= 2\cosh(\frac{\beta}{2}\sqrt{\epsilon^2+\Delta_0^2})$ and  free energy 
$F=-k_{\rm B}T\ln Z$, we have the dielectric (or charge) susceptibility
$\chi(T,\epsilon=0)$ and specific heat $C(T,\epsilon=0)$ of the
symmetric two-level system,
\begin{align}
& \chi(T) = \frac{1}{2\Delta_0}\tanh(\beta\Delta_0/2)\label{eq:chi-alpha0}\\
& C(T) = \left( \frac{\beta\Delta_0}{2} \right)^2 \sech^2(\beta\Delta_0/2)\label{eq:spec-alpha0},
\end{align}
where $\Delta_{0}$ is the low-energy thermodynamic scale \change{and
$\beta=1/k_{\rm B}T$}. Similarly,
for $P(t)$ we have at zero temperature $P(t)=\cos(\Omega_{0} t)$ with $\Omega_{0}=\Delta_0$ being the frequency of tunneling
oscillations, which for this special case of $\alpha=0$ coincides with
the thermodynamic scale $\Delta_{0}$. At finite $\alpha$ ($U<\infty$), the frictional effects
of the environment renormalize both the dynamic scale
$\Omega_{0}\to \Omega_{r}(\alpha)$ and the thermodynamic scale
$\Delta_{0}\to \Delta_{r}(\alpha)$, such that they bifurcate from
their common value at $\alpha=0$, and an additional decay (or relaxation)
time scale \change{$1/\gamma_{r}(\alpha)$} enters the time-dependent dynamics \cite{Weiss2008}.
In general, we shall define the thermodynamic scale at finite $\alpha$, by
$T_{0}\equiv\Delta_r(\alpha)=1/2\chi(0)$, so that it coincides with $\Delta_0$ at
$\alpha=0$. This scale can also be considered as the relevant low-temperature
Kondo scale since the charge susceptibility in the IRLM corresponds
to the spin susceptibility in the equivalent AKM.  
While $\Omega_r(\alpha)$ is expected to vanish at $\alpha=1/2$, signaling
a crossover to incoherent dynamics at $\alpha>1/2$\cite{Weiss2008}, the renormalized
tunneling amplitude $\Delta_r$ vanishes only at $\alpha=1^-$, with
quantum mechanical tunneling absent at $\alpha>1$ [i.e.,
$\Delta_r(\alpha>1)=0$]. At $\alpha=1$, there is a quantum critical
point with a phase transition of the Kosterlitz-Thouless type (in the
IRLM, this occurs at
$U=U^{*}=-(2/\pi\rho)\tan[\pi(\sqrt{2}-1)/2]\approx -0.969$), which
has been established by using the equivalence of the Ohmic two-state
system to the AKM\cite{Bray1982,Leggett1987}. The Anderson-Yuval scaling equations for the
latter \cite{Anderson1970} also yield an estimate for the renormalized
tunneling amplitude $\Delta_r$ at finite $\alpha$, namely, 
$\Delta_r(\alpha)/\omega_c =
(\Delta_0/\omega_c)^{1/(1-\alpha)}$. Note that the latter result
begins to differ from the exact low-energy scale $T_{0}(\alpha)$ for
$\alpha\gtrsim 1/2$ \cite{Costi1996} (see also
Fig.~\ref{fig:scale-flatband} in Sec.~\ref{subsec:thermodynamics}).

\section{Methods}
\label{sec:methods}
\subsection{NRG and TDNRG}
\label{subsec:nrg+tdnrg}
We briefly outline the NRG\cite{Wilson1975} and TDNRG\cite{Anders2005} approaches and refer the reader to 
Refs.~\onlinecite{KWW1980a,Bulla2008} and \onlinecite{Anders2005,Nghiem2014a}, respectively, for further details. 
The starting point to treat the IRLM in Eq.~(\ref{eq:IRLM}) with the NRG is a separation of the many energy scales in the conduction band via a logarithmic discretization $\epsilon_{k}=\pm D,\pm D \Lambda^{-(1-z)},\pm D \Lambda^{-(2-z)},\dots$.
Following Oliveira et al. \onlinecite{Oliveira1994}, we have
introduced a parameter $z$. Averaging physical quantities over several
realizations of the band, defined by the parameter $z$, eliminates oscillations due to the logarithmic
discretization of the band. These artificial oscillations are particularly evident in physical
quantities calculated at large $\Lambda\gg 1$. In addition, this $z$-averaging procedure
also proves useful in reducing artifacts in the time-dependent TDNRG results. Applying  the Lanczos procedure with starting state defined by the local Wannier orbital
$f_{0}=\sum_{k}c_k$ to generate a new tridiagonal basis $f_{n},n=0,1,\dots$ for $H_c$ results  in the linear chain form
\begin{align} 
&H_{\rm IRLM} =\varepsilon_{d}n_{d}+V(f_{0}^{+}d+d^{+}f_{0}) +U(n_{d}-1/2)(n_{0}-1/2) \nonumber\\
& +\sum_{n=0}^{\infty}\epsilon_{n}f_{n}^{+}f_{n} +  \sum_{n=0}^{\infty} t_{n}(f_{n}^{+}f_{n+1}+ f_{n+1}^{+}f_{n})\nonumber
\end{align} 
where the hoppings $t_{n}$ decay exponentially along the chain (as $t_n\sim\Lambda^{-(n-1)/2}$, for $n\gg1$), and, the onsite energies $\epsilon_n$, which are in general finite, are zero for the particle-hole symmetric bands that we shall consider in this paper.
Truncating the Hamiltonian $H_{\rm IRLM}$ to the first $m+1$ conduction orbitals $n=0,1,\dots,m$ and denoting this by  $H_{m}$, we have the following 
recursion relation for the $H_{m}$,
\begin{align}
&H_{m+1} = H_{m} + \epsilon_{m+1}f_{m+1}^{\dagger}f_{m+1}+t_m (f_{m}^{+}f_{m+1}+ f_{m+1}^{+}f_{m}).\nonumber
\end{align}
By using this recursion relation,  we can iteratively diagonalize the
sequence of Hamiltonians $H_m, m=0,1,\dots,$ up to a maximum chain length $m=N$.
The Hilbert space grows by a factor of $2$ at each stage, since each orbital $f_n$ can be either empty or occupied. Truncation to the lowest $N_{\rm kept}$ states is required when $m>m_{0}$, where 
$m_0\approx 10$ for $N_{\rm kept}=1000$ states. We use electron number conservation in the diagonalization procedure and choose the maximum chain length in thermodynamic calculations such that
the smallest scale in $H_{N}$, given approximately by $T_N = \Lambda^{-(N-1)/2}$, is smaller than $10^{-3}T_{0}$ with $T_{0}=1/2\chi(0)$ as defined earlier. This allows thermodynamics to be calculated at all temperatures
of interest, i.e., for $10^{-3}T_{0}\leq T \leq 2D$ where $2D=2$ is
the bandwidth (in practice we also ensure that $10^{3}T_0 \ll 2D$ so
that a part of the universal high-temperature asymptotic behavior in the temperature range $D\gg 10^{3}T_{0}\gg T \gg T_{0}$ is also captured, see Sec.~\ref{subsec:thermodynamics} for
details of how this is achieved). 
We average thermodynamic quantities, calculated by the conventional approach \cite{KWW1980a,Gonzalez-Buxton1998,Campo2005}, over several z-values, specified by  $z_{i}=(2i-1)/2n_z, i=1,n_z$ and use
$n_z=8$ and $\Lambda=4$. The thermodynamic quantities of interest are
the impurity contributions in which the host contribution is
subtracted out. Thus, for the specific heat we have $C_{\rm
  imp}(T)=C_{\rm tot}(T)-C_0(T)$ where $C_{\rm tot}(T)=k_{\rm B}\beta^2\langle\left(H_{\rm IRLM}-\langle H_{\rm IRLM}\rangle\right)^2\rangle$ 
and  $C_0(T)=k_{\rm B}\beta^2\langle \left(H_{\rm bath}-\langle H_{\rm
    bath}\rangle\right)^2\rangle$ are the specific heats of the total
system and that of the host, respectively. Similarly, for the
susceptibility, $\chi_{\rm imp}(T)$ , we have
$\chi_{\rm imp}(T)=\chi_{\rm tot}(T)-\chi_{0}(T)$, where $\chi_{\rm tot}(T)=\beta
\langle(\hat{N}-\langle\hat{N}\rangle)^{2}\rangle$ and $\chi_{0}(T)=\beta
\langle(\hat{N}_{0}-\langle\hat{N}_{0}\rangle)^{2}\rangle$ are the
charge susceptibilities of the total system and that of the host,
respectively, and $\hat{N}$ and $\hat{N}_{0}$ the corresponding total electron
number operators. The low-energy Kondo scale $T_0=1/2\chi(T=0)$ is extracted as in Eq.~(\ref{eq:chi-alpha0}) from the local susceptibility
 $\chi(T)=-\partial n_{d}/\partial \veps_d|_{\veps_d\to 0}$. Note also
 that it is well known from the equivalence between the IRLM and the
 AKM, that the two susceptibilities $\chi(T)$ and
 $\chi_{\rm imp}(T)$ differ only by a factor $\alpha$, i.e., $\chi_{\rm
   imp}(T)=\alpha\chi(T)$ \cite{Vigman1978,Costi1999}. Thus, while we
 calculate $\chi_{\rm imp}(T)$, we shall show results for $\chi_{\rm
   imp}(T)/\chi_{\rm imp}(0)$ which are equal to the susceptibility of
 interest in the Ohmic two-state system $\chi(T)/\chi(0)$.

In TDNRG, we are interested in the dynamics of a local observable $\hat{O}$
following a quantum quench 
 in which one or more system parameters of $H=H_{\rm IRLM}$ change suddenly at 
 $t=0$. Thus, the time dependence of $H$ is described by 
$H(t)=\theta(-t)H^{i} + \theta(t)H^{f}$, with  $H^{i}$ and $H^{f}$ being time-independent 
initial- ($t<0$) and final-state ($t>0$) Hamiltonians, respectively
\cite{Costi1997}. 
The time evolution of $\hat{O}$ at $t>0$ is then given by $O(t)={\rm Tr}\left[\rho(t)\hat {O}\right]$ 
where $\rho(t)=e^{-iH^{f}t}\rho\,e^{iH^{f}t}$ is the time-evolved density matrix and 
$\rho=e^{-\beta H^{i}}/{\rm Tr}[\rho]$ is the  equilibrium density matrix of the initial state at inverse temperature $\beta$. 
As shown in Ref.~\onlinecite{Anders2005}, $O(t)$ can be evaluated by
making use of the complete basis set of discarded states to give
\begin{align}
O(t)  &=\sum_{m=m_0}^N \sum_{rs\notin KK'}\rho^{i\to f}_{sr}(m) e^{-i(E^m_s-E^m_{r}) t} O^m_{rs} \label{eq:localOt},
\end{align}
in which $r$ and $s$ may not both be kept states, $O^m_{rs}={_f}\langle lem|\hat{O}|rem\rangle_f$ are the final-state matrix elements of $\hat{O}$,
which are independent of the environment variable $e$ labeling the complete set of discarded states $|lem\rangle$ (see Ref.~\onlinecite{Anders2006} for details). In deriving  the above, use has been made of the NRG approximation 
\begin{align}
&H^{f}|rem\rangle\approx H^{f}_{m}|rem\rangle=E^m_r|rem\rangle,\label{eq:nrg-approx}
\end{align}
and $\rho^{i\to f}_{sr}(m)=\sum_e{_f}\langle sem|\rho | rem\rangle_f$ 
represents the reduced density matrix of the initial state projected
onto the basis of final states and termed the 
{\em projected density matrix}. The latter has been evaluated for the 
special choice of a density matrix defined on the longest Wilson chain 
\begin{align}
&\rho=\sum_{l}|lN\rangle{_i} \frac{e^{-\beta E_l^N}}{Z_N}{_i}\langle lN|, \label{eq:rho-Anders}
\end{align}
with $Z_{N}=\sum_{l}e^{-\beta E_{l}^{N}}$, in which only the discarded 
states of the last NRG iteration enter \cite{Anders2005,Anders2006}. More recently, the projected 
density matrix has been evaluated for a general initial density 
matrix, given by the full density matrix \cite{Weichselbaum2007,Weichselbaum2012} of the initial state, 
in Refs.~\onlinecite{Nghiem2014a,Nghiem2014b}. This approach allows calculations to be carried out at 
zero or arbitrary temperature and is the approach that we use to obtain the results in
this paper. It can be shown that the TDNRG is exact in the short-time 
limit, in the sense that $O(t\to 0^{+})$ recovers the exact thermodynamic value 
$O^{i}={\rm Tr}[\rho O]$ in the initial state\cite{Nghiem2014a}. The TDNRG
remains stable and can be used to simulate to infinite times, however,
the long-time limit of observables suffers from an error of typically a few
$\%$ \cite{Anders2006,Nghiem2014a}. For the quantity of interest to us in this paper, $P(t)$, 
the absolute error in $P(t\to\infty)$ varies from
approximately $10^{-6}$ at $\alpha=0.001\ll
1$ to approximately $0.07$ at $\alpha=0.9$ (see Table~\ref{Table2}). In addition to this error,
TDNRG exhibits ``noise'' at long times $t\gtrsim
1/\gamma_{r}(\alpha)$, where, for the IRLM, $1/\gamma_{r}(\alpha)$ is the relaxation time defining the decay of $P(t)$. 
This noise can be significantly reduced by the
$z$-averaging procedure, with typically $n_z=32$ \cite{Anders2006,Nghiem2014a}. Note, that the use
of $z$-averaging here is different to its use in thermodynamic
calculations. In the latter, the aim is not to eliminate noise, but to
eliminate discretization induced oscillations in physical quantities
which occur when using a large $\Lambda$. We do not use any damping 
for the time-dependent factors  $e^{-i(E^m_s-E^m_{r}) t}$ entering 
Eq.~(\ref{eq:localOt}).

For the calculations of  $P(t)$ in this paper,  we used the following
quench protocol. The initial-state Hamiltonian $H_{i}$ is given by the
IRLM with finite hybridization $V_{i}=V$, a fixed Coulomb interaction
$U_{i}=U$, and a local level $\veps_{d,i}/\Gamma \ll -1$ such that the level is
singly occupied. This corresponds to an initial-state preparation of the Ohmic two-state 
system in the state $\sigma_z = 2(n_{d}-1/2)=+1$, with the
oscillators relaxed with respect to this state of the two-level system. The final-state 
Hamiltonian $H_{f}$ is again given the IRLM with the same $V_{f}=V_{i}=V$ and $U_{f}=U_{i}=U$
as for $H_{i}$, but with level position $\veps_{d,f} = 0$
corresponding to an unbiased Ohmic two-state. This protocol is used
also in the TD-DMRG and FRG calculations. As we wish to compare with these methods, we
shall also restrict our TDNRG calculations for $P(t)$ to zero temperature.

\subsection{DMRG and TD-DMRG}
\label{subsec:td-dmrg}
We use the numerically exact DMRG scheme formulated in matrix product states (MPS) to benchmark the results obtained within NRG at arbitrary spin-boson coupling $\alpha$ (or $U$ in the language of the IRLM). A detailed introduction to the DMRG using MPS can be found in one of the many good reviews (e.g., Refs.\onlinecite{Schollwoeck2005,Schollwoeck2011}). Using MPS means that we rewrite any quantum state via
\begin{equation}
\begin{split}
\left|\Psi\right\rangle&=\sum\limits_{n_1,\dots,n_N}c_{n_1,\dots,n_L}\left|n_1,\dots,n_N\right\rangle\\&=\sum\limits_{n_1,\dots,n_N}A^{n_1}\dots A^{n_L}\left|n_1,\dots,n_N\right\rangle
\end{split}
\end{equation} 
as a product of (local) matrices $A^{n_i}$, where $n_i$ is the local degree of freedom (such as the occupancy of the $i$-th site). Such a decomposition is always possible, but the size of the matrices $A$ grows exponentially in system size. Fortunately,  one can usually capture all relevant physics approximating the matrices $A$ by a singular value decomposition and keeping only the most relevant singular values. The square of the singular values discarded is denoted by $\chi$ in the following and we always choose $\chi\approx 10^{-7}\dots 10^{-9}$ such that the numerics are converged with respect to this numerical parameter. 
By this procedure, numerically exact results are obtained while maintaining feasible sizes for the matrices $A$.
 
For non-translation-invariant systems DMRG can be applied to a finite system only and thus to address the IRLM we use a lead of finite length $L$. The total size of the system (reservoir plus dot) is thus $N=L+1$. By performing separate calculations for different $L$, we verified that the considered $L$ are large enough such that finite-size effects can be disregarded (for the time scales considered). It is important to stop the calculation before recurrence effects (information of the end of the chain has traveled through the lead to the dot) set in. Furthermore, for simplicity we concentrate on a semi-elliptic density of states in the reservoir leading to 
\begin{align} 
&H_{\rm IRLM} =\varepsilon_{d}n_{d}+V(f_{0}^{+}d+d^{+}f_{0}) +U(n_{d}-1/2)(n_{0}-1/2) \nonumber\\
& +  D/2 \sum_{n=0}^{L-1} (f_{n}^{+}f_{n+1}+ f_{n+1}^{+}f_{n}),\nonumber
\end{align} 
which is  convenient for standard DMRG implementations as only nearest-neighbor terms have to be treated. Here, the bandwidth is $2D$. We concentrate on the same quench protocol as described in the previous section \ref{subsec:nrg+tdnrg}. To address the dynamics, we use a two-site version of an iterative ground-state algorithm (to find the initial state) as well as a symmetric fourth-order Suzuki-Trotter decomposition for the (subsequent) time evolution, as described in Chaps. 6 and 7 of Ref.~\onlinecite{Schollwoeck2011}, respectively.
 
The main idea of the iterative ground-state algorithm is to start from some initial occupancy configuration of fermions in the lead plus dot space and then repetitively sweep through the chain from left to right and right to left, optimizing the occupancy configuration of two sites at a time with respect to the total energy. For a sufficiently large number of sweeps, the energy converges and the ground state is achieved. During the sweeping process, the dimensions of the matrices $A$ increase and thus have to be truncated using a singular value decomposition (as described above). 

After preparing the ground state $\left |\Psi \right\rangle$, we perform a real-time evolution to obtain $\left |\Psi \right\rangle(t=n\tau)$ by applying $n$ times a Trotter decomposed version of $e^{-iH \tau}$. To achieve this, we separate even and odd sites $H=\sum_{i  {\rm \; odd}}h_i+\sum_{i {\rm \; even}}h_i$ and approximate $e^{-iH\tau}\approx U(\tau_1)U(\tau_2)U(\tau_3)U(\tau_2)U(\tau_1)$,
where
\begin{align}
 U(\tau_j)&=e^{-i\sum_{i  {\rm \; odd}}h_i \tau_j/2}e^{-i\sum_{i  {\rm \; even}}h_i \tau_j}e^{-i\sum_{i  {\rm \; odd}}h_i \tau_j/2},\\
 \tau_1&=\tau_2=\frac{1}{4-4^{1/3}}\tau,\;\;\;\;\;\;\;\;\tau_3=\tau-2\tau_1-2\tau_2.
\end{align}
Decreasing $\tau$ (increasing $n$) systematically improves on the
accuracy of the Trotter decomposition. During the time evolution, the dimensions of the matrices $A$ increase and the need for approximating them via the singular value decomposition and disregarding the smallest singular values arises. This is done efficiently after each applied $U(\tau_j)$. Eventually, at given $\chi$ the dimensions of the matrices $A$ increase beyond the numerically feasible level and the simulation has to be stopped.

For each plot we checked that the numerical parameters were chosen such that further increasing the accuracy of the DMRG does not alter the curves (on the scales shown). As we have to perform a time evolution using a Trotter decomposition, we cannot access the deep scaling limit of the IRLM, simply because decreasing $V/D$ requires to resolve larger times $t$ (and consequently a larger number of time steps $n$ and also larger reservoir size $L$ to avoid recurrence). Thus, eventually when decreasing $V/D$, one fails to perform the numerical calculation due to resource limitations. Nevertheless, in the following, values of $V/D$ as small as $V/D\approx 0.1$ can easily be addressed.

\subsection{FRG}
\label{subsec:frg}
In addition to the NRG and DMRG, we use the FRG approach to determine the time evolution of the IRLM. Within the FRG a truly infinite system can be tackled by using standard projection techniques \cite{Taylor1972}. With this the influence of the infinite noninteracting reservoir onto the dot system is incorporated exactly. Later, we will treat some of the reservoir degrees of freedom explicitly and only project out the rest, to efficiently model different reservoir density of states. Therefore, in the following we do not focus on a single level, but an extended interacting dot geometry coupled to (structureless) reservoirs. Again, we only summarize the main ideas as well as extensions needed and refer the reader to Ref. \onlinecite{Kennes2012a} for additional technical details. We employ the framework of the Keldysh Green's functions $G$ to derive an exact infinite hierarchy of flow equations of the form
\begin{equation}
\partial_\Lambda \gamma_m^\Lambda=f_m(G^\Lambda,\gamma_{m+1}^\Lambda,\gamma_m^\Lambda,\gamma_{m-1}^\Lambda,\dots)
\end{equation}
where $\gamma_m$ are the irreducible vertex functions.

The \change{$\Lambda$-dependency} is introduced by an auxiliary cutoff in the noninteracting Green function
\begin{equation}
G_0\to G_0^\Lambda,\;\;\;\;G^{\Lambda\to \infty}=0,\;\;\;\;G_0^{\Lambda\to 0}=G_0,
\end{equation} 
which leads to a successive incorporation of energy degrees of freedom  from high to low.
We concentrate on the so-called hybridization cutoff, which consists in coupling one additional reservoir with infinite temperature to each level of the (extended) quantum dot. Each of these auxiliary reservoirs couples to the corresponding dot level via a hybridization $\Lambda$. Starting at $\Lambda\to \infty$, where all degrees of freedom are cut out ($G^{\Lambda\to \infty}=0$), we integrate to $\Lambda=0$, where the artificial cutoff is removed and the physical situation is restored. The infinite hierarchy of flow equation, being an exact reformulation of the quantum many-body problem, can not be solved in its entity. Thus,  one needs to truncate it to a certain order. Here, we employ the first-order truncation scheme. In turn,  the results are controlled up to leading-order in the interaction $U$, but contain power-law resummations superior to a plain perturbative approach.  Within this lowest-order truncation scheme, the flow of the Keldysh self-energy vanishes and we need to determine the retarded self-energy contribution only. As a consequence, the interacting system can be interpreted as a noninteracting one with renormalized time-dependent parameters. For the IRLM, the corresponding flow equation takes the form 
\begin{align}
\partial_\Lambda \Sigma^{\rm ret}_{kl}(t',t)=-i\sum_{ijkl}\frac{U_{ijkl}(t)}{2}\delta(t'-t)
S_{lj}^{\rm K}(t,t).
\label{flowcomplete}
\end{align}   
with 
\begin{align}
  S^{\rm K} 
   = - i  G^{\rm ret} G^{\rm K} + i  G^{\rm K} G^{\rm adv} 
  \label{eq:SK}
\end{align}
for the used infinite-temperature hybridization cutoff. $U_{ijkl}$ denotes the antisymmetrized two-particle interaction, where the indices label single -particle levels. Equation \eqref{eq:SK} involves the real-time representation of the Dyson equations
\begin{align}
  G^{\rm ret}(t,t') &= G^{0, {\rm ret}}(t,t') + \left[G^{{\rm ret}} \Sigma^{\rm ret} G^{0,
    {\rm ret}}\right](t,t'), \label{eins}
  \\
  G^{\rm K}(t,t') &= - i G^{\rm ret}(t,0) (1 - 2 \bar n)
  G^{\rm adv}(0,t') 
 \nonumber \\ 
 & \quad \qquad + [G^{\rm ret} (\Sigma^{\rm K}_{\rm lead} +
  \Sigma^{\rm K}) G^{{\rm adv}}](t,t').
  \label{eq:GK_contrib}
\end{align}
One can now employ the ideas of \change{Ref.~\onlinecite{Kennes2012a}} to solve these in a very efficient and numerically exact fashion. To solve the remaining differential flow equation~\eqref{flowcomplete}, we employ a standard Runge-Kutta procedure. The relative and absolute tolerance of this Runge-Kutta integrator were chosen to be $10^{-6}$ and $10^{-8}$, respectively. An advantage of the FRG compared to the TDNRG or TD-DMRG methods used in this paper, is that results can be obtained with far less computational effort. Observables, such as the occupancy and thus $P(t)$, can be deduced from the Green's functions via
\begin{equation}
  \bar n_i(t) = \frac{1}{2} - \frac{i}{2} G^{\rm K}_{ii}(t,t).
\label{occu}
\end{equation}
This concludes our short and general summary of the FRG approach to transient dynamics. 

Within the approach described in Ref.~\onlinecite{Kennes2012a} one can only capture reservoirs in the scaling limit. To make contact with the NRG and DMRG, we extend the formalism of Ref.~\onlinecite{Kennes2012a} to reservoirs with an arbitrary density of states. To achieve this we keep an increasing number of reservoir sites explicit in our calculation effectively treating them as part of the dot structure. For the sites kept explicitly we use a linear geometry, where the one end couples to the dot  and the other end couples to  a structureless reservoir modeling all additional reservoir sites. The hopping within the explicitly treated  part of the reservoir can then be chosen such that the local density of states of the only site connected to the dot has the desired form. As this boundary density of states is the only relevant quantity for local dot observable, this method becomes numerically exact upon increasing the explicitly treated reservoir sites (and with them the number of optimization parameters to fit the density of states).  In practice, we find that approximately only  $20$ sites suffice to converge this process.  

\section{Results}
\label{sec:results}
\subsection{Thermodynamics}
\label{subsec:thermodynamics}
\begin{figure}[t]
  \includegraphics[width=\columnwidth]{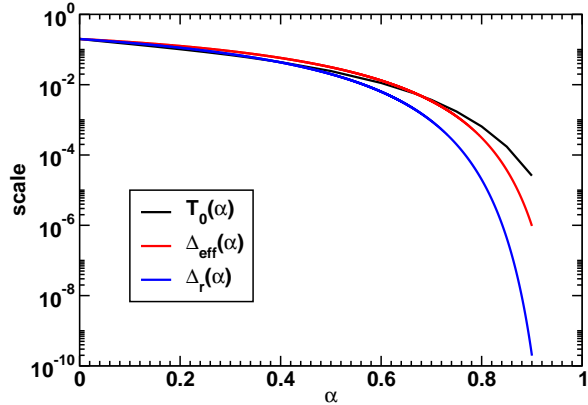}
  \caption 
  {The susceptibility scale $T_0(\alpha)=1/2\chi(0)$ vs $\alpha$ for
    the IRLM using $V=0.1D$ and a constant density of states $\rho(\omega)=1/2D$ 
    with $D=1$. Also shown are  $\Delta_{\rm  
      eff}(\alpha)$, and $\Delta_{\rm  r}(\alpha)$ vs $\alpha$. 
    NRG parameters used to  
    calculate $T_{0}(\alpha)$:  $\Lambda=4$, $n_z=8$, and $N_{\rm kept}=1000$.  
  }
  \label{fig:scale-flatband}
\end{figure}
\begin{figure}[t]
  \includegraphics[width=\columnwidth]{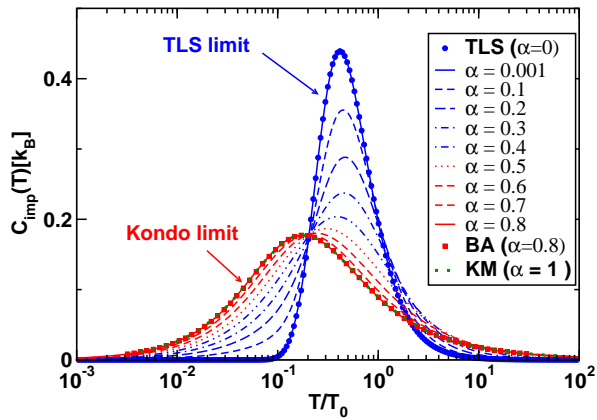}
  \caption 
{ 
Evolution of the impurity contribution to the specific heat $C_{\rm imp}$ with increasing dissipation strength $\alpha$, from weak ($\alpha<1/2$) to strong ($\alpha>1/2$)  
dissipation. Lines: results for the Ohmic two-state system obtained via the IRLM using NRG. Also shown (symbols) are the isolated two-level system result ($\alpha=0$) using  
Eq.~(\ref{eq:spec-alpha0}) and the Bethe ansatz result (via the AKM) for 
$\alpha = 4/5$ \cite{Costi1999}. KM: The universal specific heat curve for 
the isotropic Kondo model ($\alpha\to 1^{-}$) with its $T_{0}$
adjusted so that this curve coincides with the $\alpha=0.8$
curve\footnote{This adjustment of $T_{0}$ is done in order to show 
  that the $\alpha=0.8$ curve, while still not identical to the 
  isotropic Kondo model curve, nevertheless lies very close to it. 
  Without such an adjustment, the isotropic Kondo model curve would 
  lie slightly to the left.} (from Ref.~\onlinecite{Desgranges1982}). 
NRG parameters as in Fig.~\ref{fig:scale-flatband}. 
}
\label{fig:specific-heats}
\end{figure}
For the thermodynamics of the Ohmic two-state system, we are primarily
interested in universal results for specific heats and
susceptibilities in the temperature range $10^{-3}T_{0}\leq T \leq
10^3 T_0\ll D$  with $T_0$ chosen sufficiently small so that non-universal effects coming from the 
finite bandwidth $2D=2$ are minimized. For given $\alpha$
(corresponding to a given $U$ in the IRLM), this
requires choosing $V$ in the IRLM sufficiently small in order that the
highest temperatures of interest $T_{\rm max}=10^3T_{0}$ are still
much smaller than the half bandwidth $D$, i.e., $T_{\rm max}\ll D$.
Since $T_{0}$ is {\em a priori} unknown, this poses the problem of how to choose
an appropriate hybridization $V=\Delta_{0}/2$ for a given
$\alpha$.  We proceeded as follows: set $T_0(\alpha) \approx
\Delta_{\rm eff}(\alpha) =10^{-10}$ and solve for $V$. The scale
$\Delta_{\rm eff}(\alpha)$ is related to the scaling result for the renormalized tunneling amplitude 
of the Ohmic two-state system $\Delta_{\rm r}(\alpha)=\omega_c \left(\Delta_0/\omega_c\right)^{1/(1-\alpha)}$
via \cite{Weiss2008}
\begin{align}
  &\Delta_{\rm
    eff}(\alpha)=\left(\Gamma(1-2\alpha)\cos(\pi\alpha)\right)^{1/2(1-\alpha)}\Delta_{\rm
    r}(\alpha),\label{eq:delta_eff}
\end{align}
where $\Gamma$ is the $\Gamma$-function. If need be, a smaller $\Delta_{\rm
  eff}(\alpha)<10^{-10}$ can be chosen, but this was not necessary for
$\alpha\leq 0.9$. We could also have estimated a value
for $V$ by setting $T_{0}\approx \Delta_{\rm r}(\alpha)=10^{-10}$, however, the latter scale
deviates considerably more from $T_{0}(\alpha)$  than $\Delta_{\rm eff}(\alpha)$ in the regime
$\alpha>1/2$, being considerably
smaller, and hence would have resulted in unnecessarily small values of
$V$. This is illustrated in Fig.~\ref{fig:scale-flatband},
which shows the $\alpha$-dependence of $T_0(\alpha)$ and compares it
to that of $\Delta_{\rm eff}(\alpha)$, and $\Delta_{\rm
  r}(\alpha)$ for a fixed $V$ and constant density
of states (see also Table~\ref{Table2} for numerical values)
\footnote{We do not imply that $\Delta_{\rm eff}(\alpha)$ should be equal to
  the thermodynamic scale $T_{0}$. In fact,
  $\Delta_{\rm eff}(\alpha)$ is a scale that enters in the transient dynamics, 
  see  Sec.~\ref{subsubsec:comparison-niba} and
  Ref.~\onlinecite{Weiss2008}.
  Our use of $\Delta_{\rm eff}(\alpha)$ in the context of 
  this section is only as a means to obtain a rough estimate for an input $V$
  for our NRG calculations.}. Note that the adiabatic renormalization result for the low
energy scale, $\Delta_{\rm r}(\alpha)$, is close to the correct scale
$T_{0}(\alpha)$ only for  $\alpha\lesssim 1/2$. We have also checked that our
calculation of $T_{0}(\alpha)$ agrees with similar NRG calculations in Ref.~\onlinecite{Borda2008}.

\subsubsection{Specific heat}
\label{subsubsec:specific-heats}
\begin{figure}[t]
  \includegraphics[width=\columnwidth]{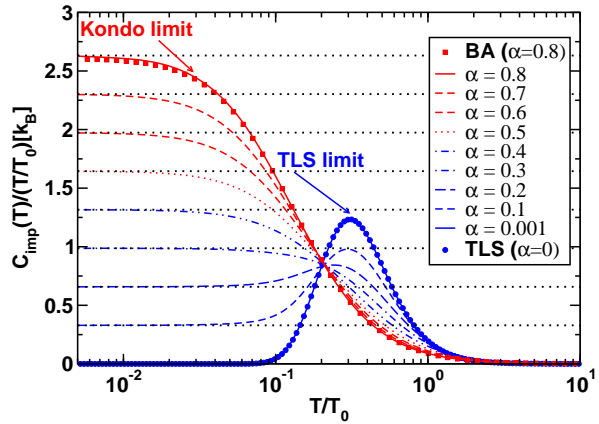}
  \caption 
{
$C_{\rm imp}/(T/T_0)$ vs $T/T_0$ for increasing dissipation strength  
$\alpha$ as in Fig.~\ref{fig:specific-heats}. Horizontal dotted lines indicate the exact Fermi liquid  
value at $T=0$, given by $\pi^2\alpha/3$ \cite{Costi1999}.  
Lines: results for the Ohmic two-state system obtained via the IRLM using NRG. Also shown (symbols) are the isolated two-level system result ($\alpha=0$) using  
Eq.~(\ref{eq:spec-alpha0}) and the Bethe ansatz result (via the AKM) for $\alpha = 4/5$ \cite{Costi1999}. NRG parameters as in Fig.~\ref{fig:scale-flatband}. 
}
\label{fig:c-over-t-fermi-liquid}
\end{figure}
We show in Fig.~\ref{fig:specific-heats} the evolution of the specific-heat curves of the Ohmic two-state system (calculated via the IRLM) 
as a function of temperature for increasing dissipation strengths ranging from very weak
dissipation ($\alpha\ll 1$) to strong dissipation strengths
($\alpha>1/2$). 
The peak position in the
specific heat shifts from $T_{\rm p}\approx 0.42T_{0}$ at $\alpha\ll
1$ to $T_{\rm p}\approx 0.17T_{0}$ at $\alpha=0.8$. Note the
approximate crossing point at $T\approx 0.2T_{0}$, a characteristic
feature in many correlated systems \cite{Vollhardt1997}.
We also show the curve for the isolated two-level system ($\alpha=0$) from
Eq.~(\ref{eq:spec-alpha0}). Although this appears to fit the
$\alpha=0.001$ curve at all temperatures, the latter, in contrast to
the former, exhibits Fermi liquid behavior at low temperatures due to
the gapless nature of the excitations in the Ohmic two-state
system. The deviations between the cases $\alpha=0$ and $\alpha\ll 1$ 
will be discussed in more detail below. We have also compared the specific
heat at $\alpha=0.8$ with that for the AKM from the numerical solution
of the thermodynamic Bethe ansatz equations of the latter
\cite{Tsvelick1983b,Costi1999}, finding good agreement at all
temperatures. Indeed, the numerical results for $C_{\rm imp}(T)$ at
asymptotically high ($T\gg T_0$) and low ($T\ll T_{0}$) temperatures
obtained from the IRLM agree with those extracted from the Bethe
ansatz for the equivalent AKM for all finite $\alpha$ \footnote{Note, that the thermodynamic
  scale in Ref.~\onlinecite{Costi1999} denoted there by $\Delta_{\rm r}$ is
  related to our present definition via $\Delta_{\rm r}=\pi T_{0}$.}:
\begin{figure}[t]
  \includegraphics[width=\columnwidth]{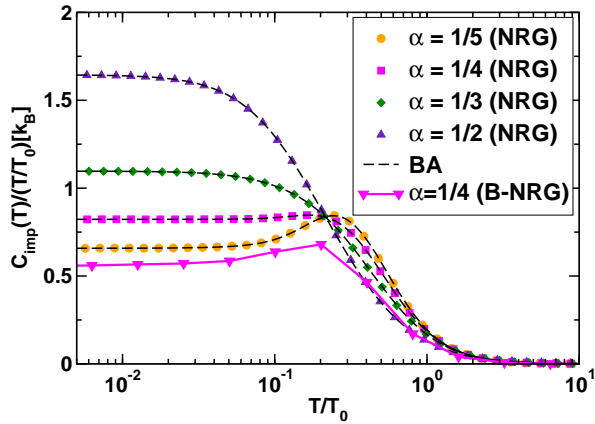}
  \caption 
{ 
$C_{\rm imp}(T)/(T/T_0)$ vs $T/T_{0}$ for the IRLM for 
$\alpha=1/5,1/4,1/3,1/2$ (symbols) compared to 
corresponding Bethe ansatz calculations (dashed lines) \cite{Costi1999}.  
NRG parameters as in Fig.~\ref{fig:scale-flatband}. For comparison, we 
also show the $\alpha=1/4$ result (B-NRG) calculated from the bosonic NRG \cite{Bulla2005}. 
}
\label{fig:c-over-t-BAcomp}
\end{figure}
\begin{align}
&C_{\rm imp}(T\gg T_{0})/k_{\rm B} \approx a\left(\frac{T_{0}}{T}\right)^{\nu_c},\label{eq:high-temp-spec}\\
&C_{\rm imp}(T\ll T_{0})/k_{\rm B}\approx \frac{\pi^{2}\alpha}{3}\frac{T}{T_{0}},\label{eq:fliq-spec}
\end{align}
with $\nu_c=2-2\alpha$ and the constant $a$ depending only on $\alpha$. The above asymptotic
behavior is also recovered in path-integral approaches\cite{Goerlich1988,Goerlich1989,Weiss2008}. 
We can test the accuracy of the Fermi liquid result by plotting
$C_{\rm imp}(T)/(T/T_{0})$ vs $T/T_{0}$ as in
Fig.~\ref{fig:c-over-t-fermi-liquid}. One sees that the result
(\ref{eq:fliq-spec}) is recovered for all finite $\alpha$ in the limit
$T/T_{0}\ll 1$ (dotted lines). Furthermore, one notices for the case
$\alpha=0.8$, that the numerical result from the IRLM is more accurate
than that from the numerical solution of the TBA equations, since the latter yield a value for 
$\lim_{T\to 0}C_{\rm  imp}(T)/(T/T_0)$ which differs from the exact Fermi
liquid result in Eq.~(\ref{eq:fliq-spec}) by $1\%$. In contrast, the IRLM result is accurate to within $0.1\%$ in this
limit. We emphasize that for general $\alpha$, Bethe ansatz
results are not readily available due to the complexity of the TBA
equations, whereas NRG calculations for the IRLM can be carried out for any
$\alpha$ and are seen to be highly accurate. In
Fig.~\ref{fig:c-over-t-BAcomp}, we compare results for $C_{\rm imp}(T)/(T/T_{0})$
obtained via the IRLM at weak dissipation $\alpha\leq 1/2$ with available Bethe ansatz
results for the AKM, finding also here excellent agreement between
these results. We also show data for $C_{\rm imp}(T)/(T/T_{0})$ at  $\alpha=1/4$ obtained
within the bosonic NRG approach to the Ohmic
spin-boson model \cite{Bulla2005}. The latter cannot be brought into 
correspondence with our IRLM curve for $\alpha=1/4$, since the finite-temperature peak in the
bosonic NRG is too large\footnote{We have adjusted the scale
  $T^{*}\sim T_0$  Ref.~\onlinecite{Bulla2005} by a factor $2$ so
  that the high-temperature asymptotes approximately match in the comparison}. Thermodynamic properties probe all
excitations of the system and might be the most difficult properties
to capture quantitatively in the bosonic NRG \cite{Bulla2005}. Increasing the number
of states and performing more refined calculations using a larger $\Lambda$ to
reduce truncation errors while implementing $z$ averaging might 
reduce the difference between these early bosonic NRG studies of the
specific heat and that calculated within the IRLM in this paper.
While the trends with $\alpha$ in $C(T)/T$ within bosonic NRG are 
qualitatively the same as in the IRLM [in particular, bosonic NRG also
describes the appearance of a finite-temperature peak in $C(T)/T$ for
$\alpha\lesssim 1/3$, see Ref.~\onlinecite{Bulla2005}], at present quantitative differences exist.

\begin{figure}[t]
  \includegraphics[width=\columnwidth]{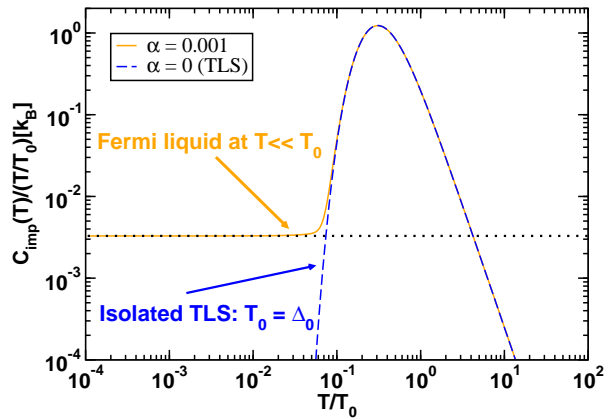}
  \caption 
{
  Specific heat divided by temperature $C_{\rm imp}(T)/(T/T_{0})$ vs
  $T/T_{0}$ of a weakly coupled Ohmic two-state system
  ($\alpha=0.001\ll 1$) compared to that of an isolated two-level system
  ($\alpha=0$). The former exhibits Fermi liquid behavior at $T\ll
  T_{0}$, and activated behavior in the range $0.1T_{0}\lesssim 
  T\lesssim 0.3T_{0}$, while the latter, having a gapped spectrum with
  $T_{0}=\Delta_0$,  exhibits activated behavior for all $T\lesssim 0.3T_{0}$. 
  Dotted line: Fermi liquid result from Eq.~(\ref{eq:fliq-spec}).
  NRG parameters as in Fig.~\ref{fig:scale-flatband}.
}
\label{fig:c-over-t-comp-TLS}
\end{figure}

We now return to discussing the aforementioned differences between the specific
heat of a weakly interacting Ohmic two-state system  (i.e., $\alpha\ll
1$) and that of an isolated two-level system (i.e., $\alpha=0$). In
Fig.~\ref{fig:c-over-t-comp-TLS},
we show on a log-log plot the comparison for $C_{\rm
  imp}(T)/(T/T_{0})$ vs $T/T_{0}$. Depicted in this way, it is evident
that the isolated two-level system provides a good description of the specific heat
of a weakly interacting Ohmic two-state system only down to temperatures
of order $0.1T_0$. Below this temperature (i.e., for $T\ll T_{0}$),
clear differences appear, with the
former exhibiting an activated behavior $C_{\rm imp(T)}(T)/(T/T_{0})
\approx k_{\rm B}(T_{0}/T)^{3}\exp(-1/(T/T_0))$, and the latter
exhibiting the Fermi liquid behavior given by
Eq.~(\ref{eq:fliq-spec}), i.e., $C_{\rm imp}(T)/(T/T_{0})\approx
k_{\rm B}\pi^2\alpha/3$ for $T\ll T_{0}$. The dotted line in Fig.~\ref{fig:c-over-t-comp-TLS} shows that
the $T\to 0$ limit of this Fermi liquid result  is indeed recovered
even for this very weakly interacting ($\alpha=0.001$) Ohmic two-state system.

We comment briefly on universality of physical quantities in the
present context. Universal results, in the rigorous sense, are obtained by taking the limit
$\Delta_{0}/\omega_{c}\to 0$ while maintaining a finite low-energy 
scale $T_{0}$. This scaling limit can be taken in analytic approaches and 
leads to a one-parameter family for physical quantities, such as specific heats 
$C_{\alpha}(T)$, which are functions of the reduced temperature 
$T/T_{0}(\alpha)$ and whose functional form is determined only by $\alpha$,
(see Refs.~\onlinecite{Tsvelick1983b,Costi1998,Costi1999}). Numerically, we always have a finite $\Delta_{0}/\omega_c=V/D$, 
resulting in non-universal corrections at high temperatures. These can, however,
be shifted to arbitrarily high temperatures by choosing a sufficiently 
small $V/D$. From the equivalence of the Ohmic two-state system to 
the AKM, the limit $\alpha\to 1^{-}$, maintaining a finite $T_{0}$
while taking  the limit $\Delta_{0}/\omega_{c}\to 0$ recovers the
universal scaling functions for the isotropic Kondo model. In this limit,
the specific heat  $C(T)=C_{\alpha=1}(T)$ acquires logarithmic
corrections at high temperatures instead of the power-law corrections
(\ref{eq:high-temp-spec}) for $\alpha<1$
\cite{Tsvelick1983b}. In Fig.~\ref{fig:specific-heats}, we show for comparison
the universal specific-heat curve of the isotropic Kondo model
$C_{\alpha=1}$ \cite{Tsvelick1983b,Desgranges1982,Desgranges1985} with
Kondo scale adjusted to match $T_{0}(\alpha=0.8)$. While the
$C_{\alpha=0.8}$ curve is strictly different from the $C_{\alpha=1}$
curve at asymptotically high temperatures, we see that, 
for the temperature range shown, the two curves are very close.
\begin{figure}[t]
  \includegraphics[width=\columnwidth]{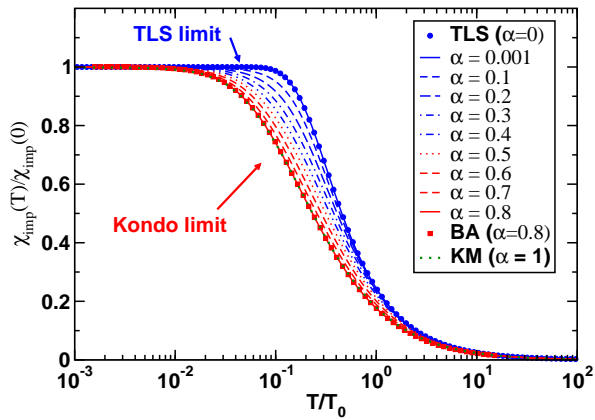}
  \caption 
{
Impurity contribution to the charge susceptibility 
$\chi_{\rm imp}(T)$ normalized to its $T=0$ value vs $T/T_{0}$ for increasing dissipation strength $\alpha$, from weak ($\alpha<1/2$) to strong ($\alpha>1/2$) 
dissipation. Lines: results for the Ohmic two-state system obtained 
via the IRLM. 
Also shown (symbols) are the isolated two-level system result ($\alpha=0$) using 
Eq.~(\ref{eq:chi-alpha0}) and the Bethe ansatz result 
for $\alpha = 4/5$ (via the equivalence to the AKM \cite{Costi1999}). 
KM: the universal susceptibility heat curve for 
the isotropic Kondo model ($\alpha\to 1^{-}$) with $T_{0}$ adjusted to
coincide with that for $\alpha=0.8$ (from Ref.~\onlinecite{Tsvelick1982}). NRG parameters as in Fig.~\ref{fig:scale-flatband}.
}
\label{fig:susceptibility}
\end{figure}
\subsubsection{Susceptibility}
\label{subsubsec:susceptibility}
The temperature dependence of the normalized impurity susceptibility $\chi_{\rm
  imp}(T)/\chi_{\rm imp}(0)$ is shown in Fig.~\ref{fig:susceptibility} 
for a range of dissipation strengths ranging from weak ($\alpha\ll 1$)
to strong ($\alpha>1/2$) dissipation. As for the specific-heats, we
see also here how the susceptibility of the isolated two-level system,
given by Eq.~(\ref{eq:chi-alpha0}), smoothly evolves with increasing
dissipation strength into the spin susceptibility of the (isotropic) Kondo model
as $\alpha\to 1^{-}$. The equivalence of the IRLM to
the AKM implies that the charge susceptibility $\chi_{\rm
  imp}$ of the
former maps onto the spin  susceptibility of the latter (since
$n_{d}-1/2$ in the IRLM maps onto $S_z$ in the AKM). Indeed, we find
that our numerical results for the IRLM at asymptotically low ($T\ll
T_{0}$) and high ($T\gg T_{0}$) temperatures correspond to those of
the AKM as extracted from the Bethe ansatz solution,
\begin{align}
&\chi_{\rm imp}(T\gg T_{0})\approx
  \frac{\chi_{\rm imp}(0)}{4}\frac{T_0}{T}\left[1-4b\left(\frac{T_{0}}{T}\right)^{\nu_{\chi}}\right],\label{eq:high-temp-chi}\\
&\chi_{\rm imp}(T\ll T_{0})\approx\chi_{\rm imp}(0)\left[1 -c \left(\frac{T}{T_{0}}\right)^2\right],\label{eq:fliq-chi}
\end{align}
with $\nu_{\chi}=\nu_c=2-2\alpha$ and constants $b, c$ depending only
on $\alpha$ (see Table~\ref{Table1} and Fig.~\ref{fig:high-T-asymptotics}).

\subsubsection{Wilson ratio}
\label{subsubsec:wilson ratio}
\begin{figure}[t]
  \includegraphics[width=\columnwidth]{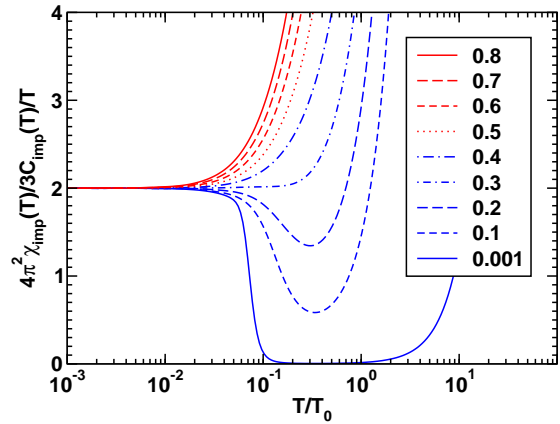}
  \caption 
{$4\pi^2\chi_{\rm imp}(T)/3C_{\rm imp}(T)/T$ vs $T/T_{0}$. This 
  tends to the Wilson ratio $R_{\rm AKM}=2$ in the limit $T\to 
  0$ for all $\alpha$ to an accuracy of less than $0.1\%$. 
 The corresponding quantity for the Ohmic 
  two-state system, $4\pi^2\chi(T)/3C_{\rm imp}(T)/T$ with 
  $\chi(T)=\chi_{\rm imp}(T)/\alpha$ yields a Wilson ratio of 
  $2/\alpha$ in the limit $T\to 0$.  
}
\label{fig:wilson-ratio}
\end{figure}
We can define a Wilson ratio for the Ohmic two-state system in terms
of $\chi(T)=\chi_{\rm imp}(T)/\alpha$ and $C_{\rm imp}(T)$
via $R_{\rm sb}=\lim_{T\to 0}4\pi^2\chi(T)/3C_{\rm imp}(T)/T$ in analogy to the
definition of this quantity for the Kondo and Anderson models
\cite{Hewson1997}. This yields, upon using Eqs.~(\ref{eq:fliq-spec}) and
(\ref{eq:fliq-chi}), the value $2/\alpha$. Note that this
equals the known value of $2$ in the isotropic Kondo limit
$\alpha\to 1^{-}$ when the susceptibilities $\chi(T)$ and $\chi_{\rm
  imp}$ become equal. Equivalently, one may define a Wilson ratio for
the AKM, by using the relevant susceptibility for the
latter, $\chi_{\rm imp}$, i.e., $R_{\rm AKM}=\lim_{T\to 0}4\pi^2
\chi_{\rm imp}(T)/3C_{\rm imp}(T)/T$. The latter is then exactly $2$ for all
$\alpha$, i.e, for all anisotropies in the AKM \cite{Tsvelick1983b}.
This is demonstrated numerically in Fig.~\ref{fig:wilson-ratio} where we show the temperature dependence of 
the quantity $4\pi^2\chi_{\rm imp}(T)/3C_{\rm imp}(T)/T$ for a range 
of dissipation strengths. The numerical values for $R_{\rm
  AKM}$ lie within $0.2\%$ of the exact value for all $\alpha$ (see
Table~\ref{Table1}). Note the different approach of  $4\pi^2\chi_{\rm imp}(T)/3C_{\rm imp}(T)/T$
to the universal  $T=0$ value of $2$ upon decreasing temperature below
$T_{0}$ for small- and large-$\alpha$ cases. The 
smallness of the above quantity in the range  $0.1\leq T/T_{0}\leq 1$ at small 
$\alpha\lesssim 0.3$ is due to the appearance of a peak in $C_{\rm 
  imp}(T)/T$ at weak dissipation, reflecting the onset of activated like 
behavior in this limit (see Fig.~\ref{fig:c-over-t-fermi-liquid}). 
\begin{table}
\begin{ruledtabular}
\begin{tabular}{dddd}
\multicolumn{1}{c}{$\alpha$} & 
\multicolumn{1}{c}{$\nu_{c}$} & 
\multicolumn{1}{c}{$\nu_{\chi}$} & 
\multicolumn{1}{c}{$R_{\rm AKM}$}\\
\colrule
0.001 & 2.00 & 2.03 &  2.00017 \\
0.1 & 1.79 &  1.78 & 2.0002\\
0.2 & 1.59  &  1.61 & 2.0019 \\
0.3 & 1.39 &  1.39 & 2.00079\\
0.4 & 1.20 & 1.20 &  2.00018\\
0.5 & 1.00 &  0.99 & 2.00002\\ 
0.6 & 0.80 &  0.80 & 2.00009\\ 
0.7 & 0.61 &  0. 61& 2.0002\\
0.8 & 0.45\footnote{The stronger deviation from the expected value 
      $2-2\alpha=0.4$ for this case indicates that $V/D$ needs to be 
      reduced further in order to access the leading high-temperature 
      correction.} & 0.41 & 2.0025\\ 
\hline 
\end{tabular}
\end{ruledtabular}
\caption{Numerical estimates of $\nu_{c}$ and $\nu_{\chi}$ entering the high 
  temperature specific heat and susceptibilities in Eq.~(\ref{eq:high-temp-spec}) and 
  (\ref{eq:high-temp-chi}).  Wilson ratio $R_{\rm AKM}$. 
}
\label{Table1}
\end{table}

\subsection{Transient dynamics}
\label{subsec:transient-dynamics}
In this section, we present TDNRG results for
$P(t)=\langle\sigma_{z}(t)\rangle$ of the symmetric Ohmic two-state
system at short [$t\ll 1/\Delta_{\rm eff}(\alpha)$] to intermediate
[$t\sim 1/\Delta_{\rm eff}(\alpha)$] time scales for the whole range of
dissipation strengths $0<\alpha<1$ and compare these with results from the NIBA
\cite{Leggett1987} (Sec.~\ref{subsubsec:comparison-niba}), the
TD-DMRG\cite{Cazalilla2002,Luo2003,Daley2004,White2004,Schollwoeck2011}(Sec.~\ref{subsubsec:comparison-td-dmrg}) and
the FRG \cite{Metzner2012,Kennes2013a,Kennes2013b} (Sec.~\ref{subsubsec:comparison-frg}). The methods are
complementary: TDNRG and TD-DMRG are non-perturbative in the
dissipative coupling $\alpha$ and can be used
to investigate the dynamics at both weak ($\alpha<1/2$) and strong
dissipation ($\alpha>1/2$), while the regime of validity of the NIBA
is generally believed to lie in the region $0\leq \alpha\lesssim 1/2$
(and not too long times, see below). The FRG for the
IRLM is, by construction, exact at $\alpha=1/2$ (corresponding to
$U=0$ in the IRLM), and remains quantitatively accurate in a finite interval around $\alpha=1/2$,
becoming inaccurate in the limits $\alpha\to 0$ (corresponding to $U\to\infty$
in the IRLM) and $\alpha\to 1^{-}$ (corresponding to $U\to
U^{*}=-0.969$ in the IRLM). The comparisons below shed further light
on the validity of the various approaches in different regimes and
time ranges. The accuracy of the NIBA, for example, has not been
convincingly tested against other reliable methods for general values
of $\alpha$, despite the considerable literature on the Ohmic
two-state system, and recent results revise the picture of the crossover from 
coherent to incoherent dynamics\cite{Kennes2013b,Kennes2013c},
see Sec.~\ref{subsubsec:comparison-frg} for a discussion of this.

\subsubsection{Comparison with NIBA}
\label{subsubsec:comparison-niba}
\begin{figure}[t]
  \includegraphics[width=\columnwidth]{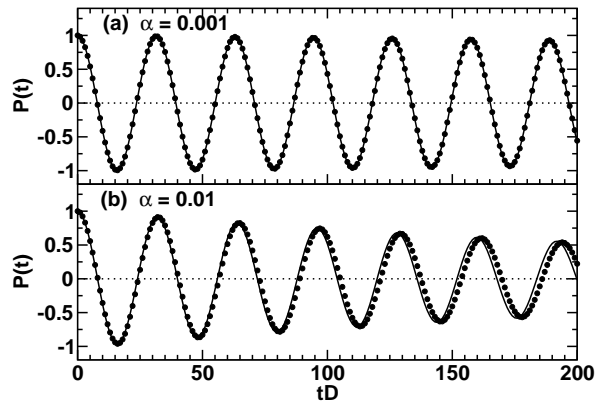}
  \caption 
{
    $P(t)$ vs $t$ in units of $1/D$ for, (a), $\alpha=0.001$, and,
    (b), $\alpha=0.01$. TDNRG (filled circles), NIBA (solid lines). 
    The TDNRG results are obtained via the IRLM with 
    a constant density of states $\rho=1/2D$, $D=1$ and 
    $V/D=\Delta_{0}/\omega_c = 0.1$. 
    NRG parameters: $\Lambda=1.6$, $N_{\rm kept}=2000$, and 
    $n_{z}=32$. The NIBA results  were obtained from Eq.~(\ref{eq:niba-expression}).
}
\label{fig:niba-alpha0p001alpha0p01}
\end{figure}
For the purpose of comparing TDNRG results with NIBA results it is
useful to first briefly summarize the content of the NIBA for $P(t)$.  At
$T=0$, the NIBA provides the following analytic expression for $P(t)$ \cite{Leggett1987,Weiss2008} 
\begin{align}
&P_{\rm NIBA}(t) = E_{2-2\alpha}\left[-(\Delta_{\rm eff}(\alpha)t)^{2-2\alpha}\right],\label{eq:niba-expression}
\end{align}
where $E_{\nu}(z)$ is the Mittag-Leffler function and $\Delta_{\rm
  eff}(\alpha)$  is defined in Eq.~(\ref{eq:delta_eff}). 
The NIBA is generally considered to be a reasonable approximation for
the short-time [$t\ll 1/\Delta_{\rm eff}(\alpha)$] to intermediate time
[$t\sim 1/\Delta_{\rm eff}(\alpha)$] dynamics of the Ohmic two-state system in
the regime $0\leq \alpha \leq 1/2$ (being exact at $\alpha=0$ and at
$\alpha=1/2$). Specifically, for $0\leq \alpha < 1/2$, Eq.~(\ref{eq:niba-expression})
predicts (for not too long times) coherent oscillations with frequency $\Omega_r(\alpha) =\cos(\pi 
\alpha/2(1-\alpha))\Delta_{\rm eff}(\alpha)\theta(1/2-\alpha)$ and decay rate 
$\gamma_r(\alpha) =\sin(\pi \alpha/2(1-\alpha))\Delta_{\rm eff}(\alpha)$. 
At $\alpha=1/2$,  the NIBA result $P_{\rm NIBA}(t)=\exp(-2\Gamma t)$
is exact (in the scaling limit of the SBM). Here, $\Gamma=\pi (\Delta_{0}/\omega_c)^2\omega_c/4=\pi\rho V^{2}$ is the
bare resonance level width of the equivalent IRLM, upon using
$\Delta_0/\omega_c=V/D$, $\omega_c=2D$
(see Appendix~\ref{appendix:equivalences} and
Table~\ref{Table3}). While the NIBA is not justified in the regime 
$1/2<\alpha<1^{-}$, see Ref.~\onlinecite{Leggett1987,Weiss2008}, 
it is nevertheless instructive to show comparisons with the NIBA 
also in this regime. For $\alpha > 1/2$, the NIBA predicts that the dynamics is incoherent,
but with $P_{\rm NIBA}(t)$ incorrectly decaying algebraically instead of 
exponentially (see below for more details).
At asymptotically short times, $t\ll 1/\Delta_{\rm eff}(\alpha)$, NIBA yields the 
behavior $P_{\rm NIBA}(t\ll 1/\Delta_{\rm eff}(\alpha))\approx 1 -
(\Delta_{\rm eff}(\alpha)t)^{2-2\alpha}/\Gamma(3-2\alpha)$ for all $\alpha$ \cite{Leggett1987,Weiss2008}.
Notice, however, that the ultrashort time scale $1/\omega_c$,
associated with the cutoff, does not explicitly enter the NIBA
expression, so the above short-time behavior persists to $t\to 0$. 
Except at $\alpha=1/2$,  a spurious incoherent contribution in Eq.~(\ref{eq:niba-expression}) dominates at
asymptotically long times $t\gg 1/\Delta_{\rm eff}(\alpha)$, yielding a
leading contribution behaving as $P(t)\sim -1/(\Delta_{\rm eff}(\alpha)
t)^{2-2\alpha}$, whereas the correct behavior is  
an overall exponential decay of $P(t)$ for all $\alpha$ (with
oscillatory terms contributing at $\alpha<1/2$) \cite{Egger1997,Lesage1998,Kashuba2013,Kennes2013b,Kennes2013c}. We
shall elaborate in more detail on this in
Sec.~\ref{subsubsec:comparison-frg} where we explain why
the long-time errors in the TDNRG, studied extensively in
Refs.~\onlinecite{Nghiem2014a,Nghiem2014b}, 
prohibit an accurate numerical investigation of the crossover from coherent to incoherent dynamics upon increasing $\alpha$.
\begin{table}
\begin{ruledtabular}
\begin{tabular}{ccccc}
\multicolumn{1}{c}{$\alpha$} & 
\multicolumn{1}{c}{$1/\Delta_{\rm eff}(\alpha)$} & 
\multicolumn{1}{c}{$1/T_{0}(\alpha)$} & 
\multicolumn{1}{c}{$\Delta_{\rm eff}(\alpha)/T_{0}(\alpha)$} & 
\multicolumn{1}{c}{$P(\infty)$} \\
\colrule
  0.001& 5.008 & 5.018 &  1.002 &0.000003 \\
  0.01 &  5.089 &  5.18  &  1.018 &0.00016 \\
  0.1 & 6.11      &  6.97  &  1.141 &0.0045\\
  0.2 & 7.915    &  9.79  &  1.237 &0.0092\\
  0.3 & 11.098  &  14.38 &  1.296 &0.0143\\
  0.4 & 17.34    & 22.82  & 1.316 &0.0186\\
  0.5 & 31.83    &  40.92  & 1.286 &0.0414\\ 
  0.6 & 75.9      &  89.2  & 1.175 &0.0407\\ 
  0.7 & 288.6    &  270.4 & 0.940 &0.049\\
  0.8 & 3232     & 1541.9  & 0.477 &0.0556\\ 
  0.9 & 1032584 & 38800.0 & 0.038 &0.066\\ 
  \hline 
\end{tabular}
\end{ruledtabular}
\caption{Numerical value of the time scale $1/\Delta_{\rm 
    eff}(\alpha)$ for different $\alpha$. Also shown 
  is the scale $1/T_{0}(\alpha)$, the ratio $\Delta_{\rm 
    eff}(\alpha)/T_{0}(\alpha)$, and 
  TDNRG results for $P(t\to\infty)=P(\infty)$ which gives the absolute 
  error in the infinite-time limit of $P(t)$.  All results are for a constant density 
  of states as in Sec.~\ref{subsubsec:comparison-niba} and the NRG 
  parameters are as in Fig.~\ref{fig:niba-alpha0p001alpha0p01}. 
}
\label{Table2}
\end{table}

We come now to the comparisons. Figure~\ref{fig:niba-alpha0p001alpha0p01} shows comparisons of TDNRG 
results for $P(t)$ (circles) at very weak dissipation strengths ($\alpha=0.001$ and $0.01$) 
with the corresponding NIBA predictions (lines). 
For $\alpha=0.001$, both the frequency and decay rate of the oscillations 
match the NIBA result up to the longest times simulated 
(approximately $8$ periods). The period of the oscillations is 
$T=2\pi/\Omega_r(\alpha)\approx 2\pi/\Delta_{0}=10\pi$. The damping of the 
oscillations in the TDNRG is only marginally larger than those in the
NIBA and is not 
visible on the scale of the plot for this particular case. For the somewhat 
stronger, but still very weak, dissipation strength of $\alpha=0.01$, the 
NIBA data match those of the TDNRG in the first two periods, but thereafter the NIBA
oscillations \change{are in advance of the} TDNRG ones, a point that we shall return to below.

\begin{figure}[t]
  \includegraphics[width=\columnwidth]{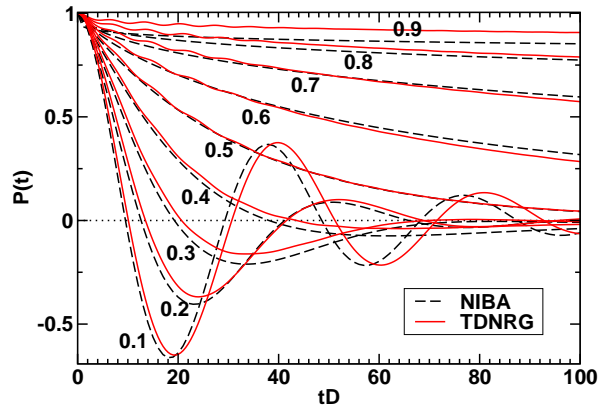}
  \caption 
{
    Comparison between TDNRG (solid lines) and NIBA (dashed lines) results for $P(t)$ for
    $0.1\le \alpha\le 0.9$. A constant density of states was
    used for the IRLM with  $V/D=\Delta_{0}/\omega_{c}=0.1$.
    NRG parameters as in Fig.~\ref{fig:niba-alpha0p001alpha0p01}.
}
\label{fig:niba-larger-alpha}
\end{figure}

With increasing $\alpha\gtrsim 0.1$, the agreement between TDNRG and NIBA first 
decreases for $\alpha$ values up to 
approximately $0.4$, with the main differences being the 
stronger damping of the oscillations in the TDNRG (particularly for
$\alpha=0.2,0.3$ and $0.4$) and a tendency of the NIBA oscillations
to \change{advance over} the TDNRG ones (particularly noticeable for
$\alpha=0.1$). Agreement then improves  
close to $\alpha=1/2$, and finally decreases again for larger 
$\alpha$, see Fig.~\ref{fig:niba-larger-alpha}. The decreasing agreement 
for $\alpha> 0.6$ is consistent with expectations about the validity 
of the NIBA in this regime, which is not expected to be quantitatively
accurate here. The decreasing agreement for increasing $\alpha$ up to
$0.4$ is more unexpected. The excellent agreement of TDNRG with 
TD-DMRG results in this region (see Fig.~\ref{fig:compare-TDNRG+TD-DMRG-semi-elliptic} in the following
section) strongly suggests that the above differences are, in fact, due to the NIBA becoming inaccurate
here. In order to further justify this statement, we proceed below to exclude
several other possibilities. 

First, we have checked that the TDNRG
results are indeed converged with respect to the number of retained states,
$N_{\rm kept}$, and with respect to the discretization 
parameter $\Lambda$, used, where smaller $\Lambda$ is known to give results closer to 
the continuum limit \cite{Anders2006,Nghiem2014a}. The results, shown in 
Figs.~\ref{fig:state-dependence-flatband} and Fig.~\ref{fig:lambda-dependence} of 
Appendix~\ref{appendix:additional-results}, indicate converged results for 
the values used, $N_{\rm kept}=2000$ and $\Lambda=1.6$. Next, we
checked that the TDNRG results for $P(t)$ are largely independent of the value
of $\Delta_{0}/\omega_c=V/D$ used (see Fig.~\ref{fig:v-dependence} of
Appendix~\ref{appendix:additional-results}). This indicates that the TDNRG
results, like the NIBA, are in the scaling limit for all interesting time
scales relative to $1/\Delta_{\rm eff}(\alpha)$. Differences between
TDNRG and NIBA at ultrashort time scales 
$t\ll 1/D\ll 1/\Delta_{\rm eff}(\alpha)$ associated with the
finite high-energy cutoff used in TDNRG calculations will be
discussed in Sec.~\ref{subsubsec:comparison-frg}. Such
cutoff-dependent 
differences, while affecting the ultrashort time dynamics,
cannot explain the deviations observed on time scales comparable to 
$1/\Delta_{\rm eff}(\alpha)$ between TDNRG and NIBA at $\alpha\sim
0.1-0.4$ [see Table~\ref{Table2} for a listing of $1/\Delta_{\rm eff}(\alpha)$].

Finally, neglecting for the moment the excellent agreement
between the TDNRG and TD-DMRG results in the following section, we
consider the possibility that the TDNRG oscillations could
be \change{in delay} over the NIBA ones due to the finite error in the
long-time limit of the former. This error increases monotonically with increasing $\alpha$ 
and is listed in Table~\ref{Table2}. The value of $P(\infty)$ should be exactly zero for an 
unbiased system, however, as discussed in detail elsewhere (see 
Ref.~\onlinecite{Nghiem2014a,Nghiem2014b}), the long-time limit of 
TDNRG observables have a finite error \footnote{This error arises from,
  (a), the use of the NRG approximation,  and, (b),  from the use of a 
  logarithmically discretized bath. The latter has a non-extensive 
  heat capacity and cannot act as a proper heat bath for 
  $\Lambda>1$. It has been argued that this prevents relaxation of the 
  system to the exact ground state of the final-state Hamiltonian 
\cite{Rosch2012}.}. 
In particular, some memory of our initial-state preparation 
$\sigma_{z}=+1$ is retained at long times, 
leading to a small positive value for 
$P(t\to\infty)$. Consequently, the oscillations at long times will 
not be exactly about the zero axis (dotted lines in 
Figs.~\ref{fig:niba-alpha0p001alpha0p01} and \ref{fig:niba-larger-alpha}), 
but somewhat shifted above this. In principle, this could lead to
\change{a delay} of the TDNRG oscillations over
the NIBA ones, but not on the time scales shown in
Fig.~\ref{fig:niba-larger-alpha}, only at longer times when $P(t)\sim
P(\infty)$. Moreover, a similar effect would have to be operative in the
TD-DMRG results in order to explain why the agreement between TDNRG
and TD-DMRG is so good, and yet they both disagree with the NIBA
results.  Since such an effect is not known in the TD-DMRG,
we conclude that the reason for the discrepancy
between NIBA and TDNRG for $0<\alpha <1/2$ is due to the simplicity of
the former approximation. The discrepancy is not insignificant, e.g.,
for $\alpha=0.3$, there are large regions in the time domain on
time scales $t\sim 1/\Delta_{\rm eff}(\alpha=0.3)$ where the relative
error is $20-30\%$. One can adjust $\Delta_{\rm eff}(\alpha)$ to fit
NIBA and TDNRG results, however, this only approximately matches the
frequencies and not the damping rates. The reason for this is that the
functional form of $P_{\rm NIBA}(t)$, which depends only on $\alpha$,
differs from the correct one. 

In Fig.~\ref{fig:niba-larger-alpha}, TDNRG results have been shown only up to 
$tD=100$ in order to facilitate comparisons with TD-DMRG results of the following 
section \footnote{For times longer than $tD=100$, the TD-DMRG calculations would require larger 
systems and correspondingly more computing resources to obtain
numerically exact results.}. This time range includes the intermediate 
time scale $1/\Delta_{\rm eff}(\alpha)$ for all $\alpha\leq 0.6$ (see
Table~\ref{Table2}), but for $\alpha\gtrsim 0.7$ this time-scale
lies outside of this range.  We therefore show in Fig.~\ref{fig:Pt-extra} of
Appendix~\ref{subsubsec:tdnrg-intermediate-times} results for $P(t)$ extending up to 
$t\sim 10/\Delta_{\rm eff}(\alpha)$ for all $\alpha$, demonstrating
that the TDNRG can access times of order $1/\Delta_{\rm 
  eff}(\alpha)$ for all $\alpha$, albeit with significant
errors at long times (tabulated in Table~\ref{Table2}).
 
Notice also a ringing effect at short times $t\gtrsim 1/D$ in the
TDNRG results in Fig.~\ref{fig:niba-larger-alpha} (particularly evident at  
large $\alpha$), which  is absent in the NIBA results. To a smaller
extent, this ringing is also present in the TD-DMRG results to be presented
below. It reflects the response of the system to a sudden
quench at $t=0$ when the only relevant time scale before the onset of
\change{final-state} correlations is the ultrashort time scale $1/D$ set by the high
energy cutoff (a time scale explicitly eliminated in the NIBA). The frequency of these
oscillations is therefore comparable to $D$. Figure~\ref{fig:wiggles} in
Appendix~\ref{subsubsec:ringing-effect} shows these oscillations and their decay in more
detail for the exactly solvable case of $\alpha=1/2$. They also occur 
in other models \cite{Anders2006,Nghiem2014a} with a hard cutoff, but
appear absent (or less pronounced) in models which use a soft cutoff
[e.g., for $J(\omega)=2\pi\alpha\omega\exp(-\omega/\omega_c)$] (see
Ref.~\onlinecite{Wang2008} for an example).

\subsubsection{Comparison with TD-DMRG}
\label{subsubsec:comparison-td-dmrg}
\begin{figure}[t]
  \includegraphics[width=\columnwidth]{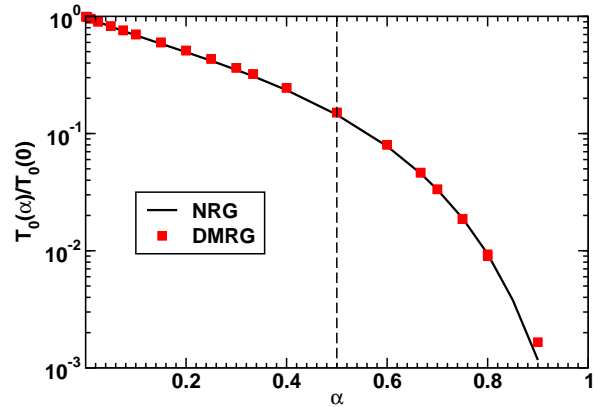}
  \caption 
  {The thermodynamic scale $T_0(\alpha)$ (normalized to its $\alpha=0$
    value) vs $\alpha$ from NRG  (solid line) and DMRG (squares). 
    Both NRG and DMRG calculations were carried out 
    for a semi-elliptic density of states $\rho(\omega)=\frac{2}{\pi  
      D^2}\sqrt{D^2-\omega^2}$ using a Wilson chain representation of 
    the conduction band.  
    For NRG $\Lambda=4$, $N_{\rm kept}=1062$ and $n_z=8$ were used, while for DMRG 
    $\Lambda=1.6$ (without $z$-averaging) was used. 
    $T_{0}=1/2\chi(T=0)$ was calculated via a numerical derivative 
    $\chi(T=0)=-\partial n_{d}/\partial\veps_d|_{\veps_d=0}\approx -\Delta  
    n_d/\Delta\veps_d$ with a  
    sufficiently small increment $\Delta\veps_d$ such that converged  
    results were obtained. 
  }
  \label{fig:t0-semi-elliptic}
\end{figure}
 The TD-DMRG calculations for the IRLM were carried out using a tight-binding 
representation of the conduction band with constant hoppings $D/2$ along the 
chain. This corresponds to using a semi-elliptic density of states 
$\rho(\omega)=\frac{2}{\pi 
  D^2}(D^2-\omega^2)^{1/2}\theta(D-|\omega|)$. By using the same choice of 
density of states in the TDNRG we can make a quantitative comparison 
of results for $P(t)$ with corresponding TD-DMRG results. Before
showing these comparisons, we first 
check that we recover the same low-energy scale $T_{0}(\alpha)=1/2\chi(0)$
with $\chi(0)=-\partial n_d/\partial\veps_d|_{\veps_d=0}$ in both 
NRG and DMRG. For this purpose, we used  a Wilson chain corresponding 
to a semi-elliptic band in both NRG and DMRG calculations. The 
results, shown in Fig.~\ref{fig:t0-semi-elliptic}, indicate very 
good agreement over the whole range $0\leq \alpha \leq 0.9$. 
In practice, the TD-DMRG calculations reported below use a tight-binding chain of
finite length $L = 200$. While the low-energy scale in equilibrium
DMRG $T_0 (\alpha, L)$ still shows prominent finite-size effects at $L=200$ (see Fig.~\ref{fig:t0-semi-elliptic-dmrg-finite-size} of 
Appendix~\ref{subsubsec:td-dmrg-l-dependence}), 
nevertheless, this is sufficient to converge the TD-DMRG results for $P(t)$
with respect to $L$ for the time scales shown in the comparisons below
($tD\leq 100$).
The reason for this is that the chosen initial  state (see
Sec.~\ref{sec:methods}) of reservoir plus impurity is influenced very
little by the finite-size of the  reservoirs with respect to the
boundary properties next to the impurity site (in contrast to the
finite size effects for the equilibrium unbiased $\varepsilon=0$ system). 
For the time evolution, the Lieb-Robinson bounds ensure that
information about the end of the chain can only travel through the 
reservoir at finite speed $c$ (here the Fermi velocity) \cite{Lieb1972}. This
defines a light cone of width $ct$ around the impurity site outside of 
which the influence of the finiteness of the reservoirs is suppressed
{\emph{exponentially}}. Once this light cone reaches the end of the
chain, prominent finite size effects set in, but at all earlier times the results
can be regarded to be in the thermodynamic limit for the local observables
of interest here. To illustrate this, we refer the reader to 
Fig.~\ref{fig:Ldependence-alpha0p25-alpha0p8} 
of Appendix~\ref{subsubsec:td-dmrg-l-dependence}.

\begin{figure}[t]
  \includegraphics[width=\columnwidth]{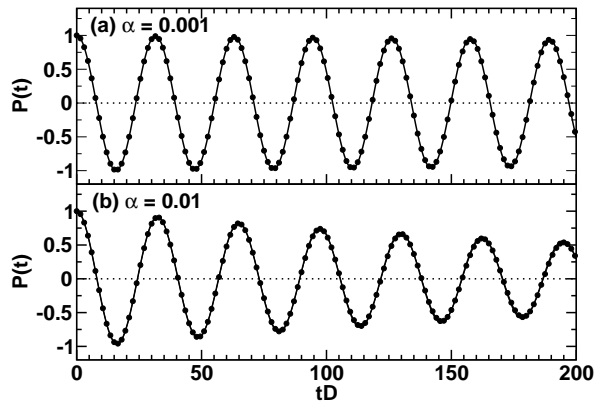}
  \caption 
{Comparison between TDNRG (filled circles) and TD-DMRG (solid lines) for, (a), $\alpha=0.001$, and,
  (b), $\alpha=0.01$. NRG parameters as in 
  Fig.~\ref{fig:niba-alpha0p001alpha0p01}.  A semi-elliptic DOS is 
  used and $V/D=\Delta_{0}/\omega_c = 0.1$. 
  NRG parameters as in Fig.~\ref{fig:niba-alpha0p001alpha0p01}. 
}
\label{fig:compare-TDNRG+TD-DMRG-semi-elliptic-small-alpha}
\end{figure}

In Fig.~\ref{fig:compare-TDNRG+TD-DMRG-semi-elliptic-small-alpha}
we compare TDNRG 
results for $P(t)$ (circles) at very weak dissipation ($\alpha=0.001$ and $\alpha=0.01$) 
with corresponding TD-DMRG results (solid lines). The agreement is
essentially perfect out to the longest times simulated, with both the frequency and damping rate
of the oscillations being almost identical. For $\alpha=0.01$, where we
previously found that the NIBA oscillations \change{advanced} somewhat \change{over} the TDNRG
ones, we here find perfect agreement between TD-DMRG and TDNRG.
\begin{figure}[t]
  \includegraphics[width=\columnwidth]{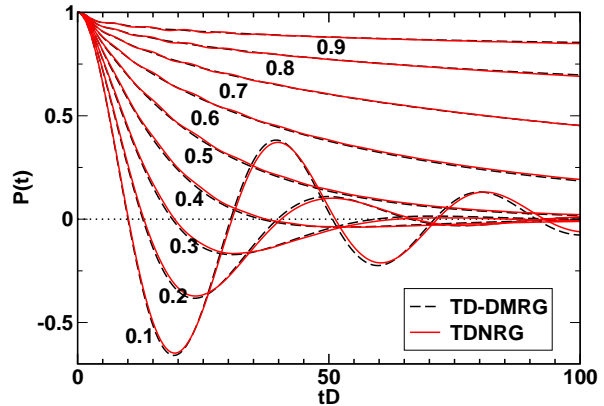}
  \caption 
{Comparison of  TDNRG (solid lines) and TD-DMRG (dashed lines) results for $P(t)$ for  
  $\alpha=0.1,\dots,0.9$.  A semi-elliptic DOS is 
  used and $V/D=\Delta_{0}/\omega_c = 0.1$. 
  NRG parameters as in Fig.~\ref{fig:niba-alpha0p001alpha0p01}. 
}
\label{fig:compare-TDNRG+TD-DMRG-semi-elliptic}
\end{figure}

With increasing $\alpha$ (see Fig.~\ref{fig:compare-TDNRG+TD-DMRG-semi-elliptic})
the agreement between TDNRG and TD-DMRG continues to be very good at
all times and up to the largest dissipation strength calculated
($\alpha=0.9$). The TDNRG results for $0.1\lesssim \alpha\lesssim 0.3
$ exhibit a marginally larger damping
than the TD-DMRG results. This could be due to the
better description of the continuum, and hence the damping, within the latter
approach on the time scales shown \footnote{
On the times $tD\leq 100$ of the comparisons, the tight-binding 
chain used in TD-DMRG with chain lengths $L=200$ gives a better 
representation of the continuum conduction band, than a Wilson 
chain with $\Lambda=1.5$, which treats high energies with less 
resolution than the low energies (due to the use of a logarithmic 
mesh of energies about the Fermi level). Conversely, for sufficiently 
long times $tD>100$, finite-size effects, discussed in 
Appendix~\ref{subsubsec:td-dmrg-l-dependence},
are more pronounced in the TD-DMRG results.}.
For $\alpha> 0.7$, and for the longest times $tD\sim 100$, one observes 
a tendency in Fig.~\ref{fig:compare-TDNRG+TD-DMRG-semi-elliptic} for the TD-DMRG 
results for $P(t)$ to lie marginally above the TDNRG ones. While the
effect is marginal, it appears to be due to an upturn of the TD-DMRG results for longer times,
$tD\gtrsim 100$ (see Fig.~\ref{fig:Ldependence-alpha0p25-alpha0p8}  in 
Appendix~\ref{subsubsec:td-dmrg-l-dependence}). For all intents and purposes, the results are converged
with respect to $L$ on the times scales of interest $tD\lesssim
100$. In contrast, the use of a Wilson chain in TDNRG, simulates an
essentially infinite (but discrete) system, and allows calculations to
be carried out to 
arbitrarily long times, without significant finite-size
effects, but with a finite error in this limit (see
Table~\ref{Table2} and Fig.~\ref{fig:Pt-extra} in
Appendix~\ref{subsubsec:tdnrg-intermediate-times}). 
Notice also the ringing at short times $t\gtrsim
1/D$ with frequency of order $D$ and discussed in the previous section.
The oscillations correlate well for all $\alpha$ between TD-DMRG and TDNRG.

As noted previously, the NIBA expression $1-P(t) \sim (\Delta_{\rm 
  eff}(\alpha)t)^{2-2\alpha}$ for $t\ll 1/\Delta_{\rm eff}(\alpha)$, being in the scaling limit, is cutoff independent, since 
the cutoff $\omega_c$ has been absorbed into the low-energy scale 
$\Delta_{\rm eff}(\alpha)$. This means, in particular, that the short-time 
exponents $2-2\alpha$ persist in the whole range $0\leq t\ll 1/\Delta_{\rm 
  eff}(\alpha)$. In contrast, both TDNRG and TD-DMRG retain information about 
the finite high-energy cutoff $D$ explicitly, thereby restricting the
time range for observing the above exponents to 
$1/D \ll t \ll 1/\Delta_{\rm eff}(\alpha)$. Since the calculations in the present 
section used $V/D=\Delta_{0}/\omega_c = 0.1$, the energy window for extracting the above 
exponents is too small for most $\alpha$ \footnote{This choice of $V$ was 
used in order to have comparable parameters as in the TD-DMRG 
calculations of Sec.~\ref{subsubsec:comparison-td-dmrg}.}. 
In Sec.~\ref{subsubsec:comparison-frg}, we shall use a much smaller 
$V$, thereby allowing the above exponents to be verified quantitatively. 

\begin{figure}[t]
  \includegraphics[width=\columnwidth]{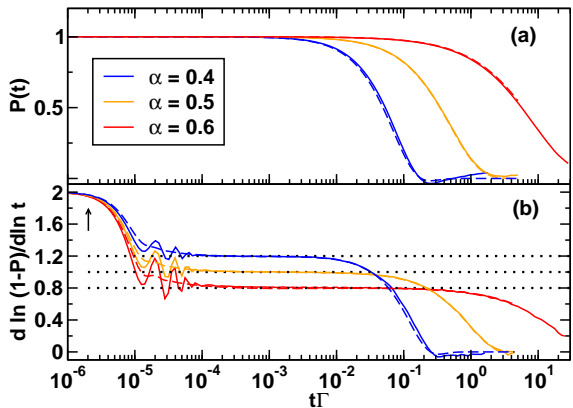}
  \caption 
{(a) Comparison of TDNRG (solid lines) and FRG (dashed lines) results 
  for $P(t)$ vs $t\Gamma$ in the region $\alpha\approx 0.5$, specifically for 
  $\alpha=0.4, 0.5$ and $0.6$. We used a semi-elliptic density of 
  states and $\Delta_{0}/\omega_c =V/D=0.001$ corresponding (for 
  $\alpha=1/2$) to a resonant level halfwidth of $\Gamma=2\times 
  10^{-6}$. In comparing results close to $\alpha=1/2$, measuring time
  in units of $1/\Gamma$ is useful, since $\gamma_{r}(1/2)=\Delta_{\rm
  eff}(1/2)=2\Gamma$. 
(b) Comparison of the logarithmic derivative $d\ln[1-P(t)]/d\ln t$ of $1-P(t)$ vs 
$t\Gamma$ from TDNRG (solid lines) and FRG (dashed 
lines) for the same $\alpha$ values as above.  Black dotted lines 
indicate the short time exponents $2(1-\alpha)$ in 
$1-P(t)\sim (\Delta_{\rm eff}(\alpha)t)^{2(1-\alpha)}$ for $1/D\ll t\ll 1/\Delta_{\rm eff}(\alpha)$. The vertical arrow 
indicates the ultrashort time scale $t=1/D$ (in units of $1/\Gamma$). 
The logarithmic derivative magnifies the ringing effect, described in
Sec.~\ref{subsubsec:comparison-niba}-\ref{subsubsec:comparison-td-dmrg}
and in Appendix~\ref{subsubsec:ringing-effect}. 
NRG parameters as in Fig.~\ref{fig:niba-alpha0p001alpha0p01}. 
}
\label{fig:compare-FRGNRG-semi-elliptic}
\end{figure}

\subsubsection{Comparison with FRG}
\label{subsubsec:comparison-frg}
Close to $\alpha=1/2$ (corresponding to the vicinity of $U=0$ in the
IRLM), FRG calculations are well controlled and
can be used to compare with TDNRG
results for $P(t)$. Figure~\ref{fig:compare-FRGNRG-semi-elliptic}~(a)
shows these comparisons for $\alpha=0.4$, $0.5$, and $0.6$ and
$V/D=\Delta_{0}/\omega_c=0.001$ where a semi-elliptic density of
states has been used for both FRG and TDNRG\footnote{For a
  semi-elliptic density of states, the mapping of 
the IRLM to the Ohmic  SBM is not rigorously valid, nevertheless 
in the scaling limit, $V/D\ll 1$, one expects the equivalence to hold on all
  time scales, except at $t\lesssim 1/D$. In extracting $\alpha$ from
  Table~\ref{Table3}, we use $\rho(0)=1/\omega_{c}=2/\pi$.}.
For $\alpha=0.5$ the agreement is particularly good at all times. FRG is
exact in this limit, whereas TDNRG entails the usual approximation
associated with neglecting high energy states, hence the small
difference at $t\gtrsim 1/\Gamma$, where the bare resonant
level halfwidth $\Gamma=\pi\rho(0)V^2$ is the relevant energy
scale in this limit. Very good agreement is also found at $\alpha=0.4$
and $0.6$. Since these calculations were for $V/D=\Delta_{0}/\omega_c=0.001$, it becomes
possible to analyze the short-time limit $1/D\ll t\ll 1/\Delta_{\rm eff}(\alpha)$ in
detail. In particular it is now possible to extract the aforementioned exponents $2-2\alpha$ in the
short time behavior of $1-P(t)\sim (\Delta_{\rm
  eff}(\alpha)t)^{2-2\alpha}$ with $t$ in the range $1/D\ll t \ll
1/\Delta_{\rm eff}(\alpha)$. That these exponents are recovered, both within
FRG and TDNRG, can be seen in
Fig.~\ref{fig:compare-FRGNRG-semi-elliptic}~(b), which shows the
logarithmic derivative $d \ln[1-P(t)]/d \ln (t)$.

Notice also, that at ultrashort time scales,  $t\ll 1/D$, there are
no bath degrees of freedom available to follow the dynamics of the two-level
system. Therefore, the dynamics is of the form $P(t)\approx
1-c_{\alpha_i} t^{2}$, similar to that of a noninteracting two-level
system in the limit $t\to 0^{+}$ with an exponent $2$ as found 
also numerically in Fig.~\ref{fig:compare-FRGNRG-semi-elliptic}~(b). 
The prefactor $c_{\alpha_i}$ depends on the dissipative coupling $\alpha_i$ in the initial state, which for the
present quench protocol (see Sec.~\ref{subsec:nrg+tdnrg}) with 
$U_i=U_f$, is equal to the dissipative coupling in the final state 
$\alpha_f=\alpha$. For quantum quenches, where $\alpha_i\neq
\alpha_f$, the ultrashort time behavior of $P(t)$ thus depends on $\alpha_i$,
whereas once system-bath correlations develop at $t\gtrsim 1/D$, a
dependence of $P(t)$ on $\alpha_f=\alpha$ results.

\begin{figure}[t]
  \includegraphics[width=\columnwidth]{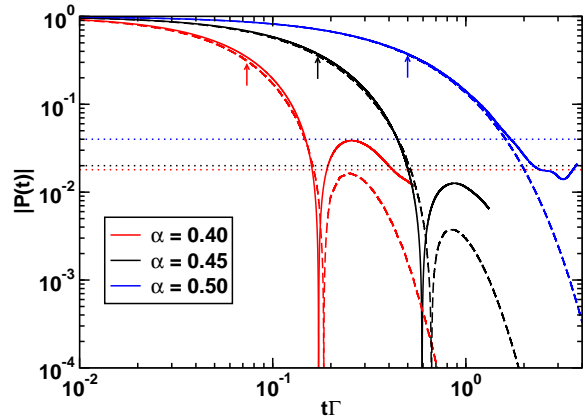}
  \caption 
  {Comparison of TDNRG (solid lines) and FRG (dashed lines) results 
    for $|P(t)|$ vs $t\Gamma$ for $\alpha=0.40, 0.45$ and $0.50$ (in 
    the partially coherent regime of Ref.~\onlinecite{Kennes2013c}) on a 
    log-log plot. Vertical arrows indicate the relaxation
    time scale $1/\gamma_{r}(\alpha)$ within the NIBA. Horizontal
    dotted lines indicate the long-time error $P(\infty)$ in the TDNRG 
    results. TDNRG results are shown up to $t \approx 7/\gamma_r(\alpha)$,
    the typical time scale beyond which $|P(t)|$ has decayed to a
    value below the long-time error$|P(\infty)|$, and thus the data
    at longer times have no significance.
    We used a semi-elliptic density of states and $\Delta_{0}/\omega_c =V/D=0.001$ corresponding (for 
    $\alpha=1/2$) to a resonant level halfwidth of $\Gamma=2\times 
    10^{-6}$. NRG parameters as in Fig.~\ref{fig:niba-alpha0p001alpha0p01}. 
}
\label{fig:crossover-region}
\end{figure}
Finally, we comment on why the TDNRG in its present formulation cannot
be used to investigate the subtle issue of the crossover from coherent
to incoherent dynamics
\cite{Leggett1987,Weiss2008}. It was recently shown\cite{Kennes2013b,Kennes2013c}, 
that this crossover is by far more complex than predicted by NIBA. In the 
latter, a single step transition from coherent dynamics 
with oscillatory behavior of $P(t)$ for all times to monotonic 
dynamics occurs at $\alpha=1/2$. However, using complementary 
RG methods, one of them being FRG, it was shown in Refs.~\onlinecite{Kennes2013b,Kennes2013c,Kashuba2013} that the 
crossover from fully coherent to incoherent dynamics occurs via an intermediate regime  
with oscillatory behavior on short to intermediate times but monotonic behavior 
at large times (see also Ref.\cite{Egger1997}). This regime was dubbed the partially coherent one.  
It would be very interesting to confirm this rich scenario using
TDNRG. However, within both the NIBA, which predicts a crossover at $\alpha_{c}=1/2$,
as well as the renormalization group approaches of
Refs.~\onlinecite{Kennes2013b,Kennes2013c}, the quality factor
of the oscillatory contributions to $P(t)$ is the same and is given by 
$Q=\Omega_{r}(\alpha)/\gamma_r(\alpha)=\cot
\frac{\pi}{2}\frac{\alpha}{1-\alpha}$. In order to investigate this
issue with the TDNRG, we require at the very least reliable results
for $P(t)$ for two periods $t=2T$ at $\alpha\approx 0.3$. 
At approximately this value of $\alpha$, the transition from fully to partially 
coherent dynamics was found. Since
$Q(0.3)\approx 1$ and $\Delta_{\rm eff}(0.3)\approx\gamma_r(0.3)$, 
we have that $P(2T)\approx P(4\pi/\Delta_{\rm eff}(0.3))\approx
P(13/\Delta_{\rm eff}(0.3))$. On time scale $t=13/\Delta_{\rm
  eff}(0.3)\gg 1/\Delta_{\rm eff}$, we can assume an exponential decay
of $P(t)$ to obtain an upper bound estimate, i.e., using $P(t)\sim
\exp(-\gamma_r(0.3)t)\approx\exp(-\Delta_{\rm eff}(0.3)t)$, we have that
$P(2T)\approx \exp(-13)\approx 2\times 10^{-6}$. Such an accuracy,  
is, however, not presently available within TDNRG, which for
$\alpha=0.3$ has a long-time error of $0.0143\gg P(2T)$ (see
Table~\ref{Table2}). This error starts to develop on a time scale of order
$1/\gamma_r(\alpha)$, thereby making the investigation of the above
issue impossible within the single-quench TDNRG formalism. In order to
illustrate this, a comparison of  TDNRG and FRG data for $|P(t)|$ in the partially coherent regime 
is shown on a log-log scale in Fig.~\ref{fig:crossover-region}.
The dips correspond to zeros of $P(t)$, the vertical arrows indicate
the NIBA time scale $1/\gamma_r(\alpha)$, and the horizontal dotted
lines indicate the long-time error $P(\infty)$ in the TDNRG\footnote{Once 
$|P[t\gg 1/\gamma_{r}(\alpha)]|<|P(\infty)|$, TDNRG results are 
dominated by the finite long-time error and have no significance, hence, we 
show TDNRG data only up to $t\approx 7/\gamma_r(\alpha)$.}. This
comparison confirms the simple argument given above, that the onset of
errors at $t\gtrsim 1/\gamma_{r}(\alpha)$ in TDNRG, studied in detail
in Refs.~\onlinecite{Nghiem2014a,Nghiem2014b} for the Anderson
impurity model, prevents, in the present
context, a detailed investigation of the crossover scenario from coherent to
incoherent dynamics of Refs.\onlinecite{Kennes2013b,Kennes2013c,Kashuba2013}. On the other
hand, the TDNRG results on time scales up to order $1/\gamma_{r}(\alpha)$
are also not inconsistent with such a picture.

\section{Conclusions}
\label{sec:conclusions}
In this paper, we investigated the thermodynamics and
transient dynamics of the Ohmic two-state system, for 
dissipation strengths ranging from weak ($\alpha<1/2$) to strong
($\alpha>1/2$), by using the equivalence of this model
to the IRLM. The IRLM, being a spinless fermionic model, is 
the simplest model that can capture the physics of the Ohmic two-state
system in the interesting parameter regime $0\leq \alpha<1$. Within an
NRG treatment of the IRLM, a larger fraction ($1/2$) of the states generated at each
NRG iteration can be retained than is possible for other equivalent
models, e.g., for the AKM (where approximately $1/4$ of the states can be
retained). Consequently, more accurate and efficient calculations of the
thermodynamics and transient dynamics of the Ohmic
two-state system can be carried out by using the IRLM.

For the thermodynamics, we showed how the universal specific heat and susceptibility curves evolve 
with increasing dissipation strength, $\alpha$, from the results for
an isolated two-level 
system at vanishingly small dissipation $\alpha\to 0$, to the results
for the isotropic Kondo model in the limit $\alpha\to 1^{-}$. The results recover in all cases
the exact Fermi liquid behavior at $T\ll T_{0}(\alpha)$ and the known
high temperature asymptotics at $T\gg T_{0}(\alpha)$. Our results via
the IRLM go beyond available Bethe ansatz 
calculations at rational values of $\alpha$ for the
equivalent AKM\cite{Costi1999}, since they can be carried out for {\em arbitrary} 
dissipation strengths $0\leq \alpha\leq 1^{-}$. Furthermore, we
found very good agreement between NRG and DMRG calculations
for the $\alpha$ dependence of $T_{0}$.

For the zero-temperature transient dynamics,  we demonstrated excellent agreement on
short to intermediate  time scales for $P(t)$ between TDNRG 
and TD-DMRG for $0\lesssim\alpha \lesssim 0.9$, and between TDNRG 
and FRG in the vicinity of $\alpha=1/2$. These comparisons (see Fig.~\ref{fig:niba-larger-alpha})
indicate that the TDNRG remains accurate, not only in limiting cases
such as $t\to 0^{+}$, where it is known to be exact, but also at
finite times $t\sim 1/\Delta_{\rm eff}(\alpha)$, and for general $\alpha$
where no exact results are available. 

Finally, our comparisons between the NIBA and TDNRG allowed us to
quantify the error in the former approximation for
a range of $\alpha$. While, it is known that NIBA is not quantitatively
correct in the incoherent regime $\alpha \gtrsim 0.5$, we found that
there are significant errors of $20-30\%$ 
also in the range $0.1\lesssim \alpha\lesssim 0.4$ 
for time scales comparable to $1/\Delta_{\rm eff}(\alpha)$. In
contrast, the NIBA agrees well with TDNRG  for $\alpha\approx 1/2$ and
for $\alpha\ll 1$ at short to intermediate times. In conclusion, our TDNRG results for $P(t)$ at general values of
$\alpha$ and for short to intermediate time scales could serve as
useful benchmarks for the development of new techniques to simulate the transient 
dynamics of spin-boson problems. 

A major problem that still needs to be overcome within the TDNRG, is that
of obtaining accurate results for transient quantities at long times,
$t\gg 1/\gamma_{r}(\alpha)$, and in the asymptotically long-time limit
$t\to\infty$\cite{Anders2006,Nghiem2014a}. Progress on
this, for example,  within a multiple-quench generalization\cite{Nghiem2014b}, or, within 
hybrid TD-DMRG/TDNRG approaches \cite{Guettge2013}, might allow
issues such as the crossover from coherent to incoherent dynamics in the Ohmic
two-state system to be investigated.

\begin{acknowledgments}
We acknowledge financial support from the Deutsche Forschungsgemeinschaft via RTG 1995 and supercomputer support by the John von Neumann institute
for Computing (J\"ulich) and the RWTH Compute Cluster. We would like to thank Ralf Bulla for
discussions and sending us data from Ref.~\onlinecite{Bulla2005} which
is used in Fig.~\ref{fig:c-over-t-BAcomp}.
\end{acknowledgments}
\appendix 
\section{Equivalence of the Ohmic spin-boson model to the AKM,
  Spin-Fermion model and the IRLM}
\label{appendix:equivalences}
\change{For completeness, we here provide details of the equivalence of the Ohmic spin-boson model to a number of fermionic 
models via bosonization \cite{Schlottmann1982,Guinea1985}.  The basic
bosonization identities that we use are summarized in
Sec.~\ref{subsec:bosonization}. For further details on bosonization,
we refer the reader to the comprehensive overview of bosonization
techniques in Ref.~\onlinecite{vonDelft1998}.} 
The procedure we follow is then to start with the AKM (Sec.~\ref{subsec:bosonized-AKM}) and apply 
unitary transformations to map this model successively onto the SBM (Sec.~\ref{subsec:equivalence-ohmic-spin-boson-model}), and the 
IRLM (Sec.~\ref{subsec:equivalence-IRLM}). On the way, we also relate the SBM to the spin-fermion model (Sec.~\ref{subsec:equivalence-spin-fermion-model}).
\subsection{Bosonization of free fermions}
\label{subsec:bosonization} 
Consider first the free fermion Hamiltonian,
\begin{equation}
H_{0}=\sum_{k\mu}\varepsilon_{k}c_{k\mu}^{\dagger}c_{k\mu}. 
\end{equation}
We take a linear dispersion 
relation for the conduction electrons, $\varepsilon_{k}=v_{\rm F}k$, and measure 
$k$ relative to the Fermi wave number $k_{\rm F}$. 
We introduce fermion fields $\psi_{\mu}(x)$ via a Fourier series,
\begin{equation}
\psi_{\mu}(x) = L^{-1/2}\sum_{k}e^{-ikx}c_{k\mu}
\end{equation}
with wave numbers $k=2\pi n/L, n=0,\pm 1,\dots$ for periodic boundary 
conditions appropriate to a finite system of length $L$. The density 
of states per spin direction is given by 
$\rho=1/2\pi v_{\rm F}$. The kinetic energy can then be written as,
\begin{equation}
H_{0}=v_{\rm F}\sum_{k,\mu} k\, c_{k\mu}^{\dagger}c_{k\mu}=
i v_{\rm F}\int_{-L/2}^{+L/2}
\psi_{\mu}^{\dagger}(x)\partial_{x}\psi_{\mu}(x)\,dx
\end{equation}
The fields $\psi_{\mu}(x)$ are expressed in terms of Hermitian bosonic
fields $\varphi(x)$ in the standard way,
\begin{equation}
\psi_{\mu}(x)=(2\pi a)^{-1/2}F_{\mu}e^{-i\varphi_{\mu}(x)},\label{eq:fermion-field}
\end{equation}
where $a$ is a cutoff, required for obtaining convergent 
momentum sums. The $F_{\mu}$ are the Klein factors required to  
ladder between states with different fermion number, 
and to ensure the correct anticommutation relations for the fermion 
fields \cite{Kotliar1996,vonDelft1998}, and
\begin{equation}
\varphi_{\mu}(x)=\phi_{\mu}(x)+\phi_{\mu}^{\dagger}(x),
\end{equation}
with the bosonic fields $\phi,\, \phi^{\dagger}$ given by
\begin{equation}
\phi_{\mu}^{\dagger}(x)=[\phi_{\mu}(x)]^{\dagger}\equiv 
-\sum_{q>0}n_{q}^{-1/2}e^{iqx}a_{q\mu}^{\dagger}e^{-aq/2}.
\end{equation}
Equation (\ref{eq:fermion-field}) only holds as an operator identity if
$a$ is sent to zero, which thus has to be done at the end of every
calculation.  In the present context taking $a \to 0$ is equivalent to considering the scaling limit which explains why the mappings discussed below only hold in this limit.
The $a_{q},\, a_{q}^{\dagger}$, defined for $q>0$, satisfy boson commutation 
relations with $n_{q}=(qL/2\pi)^{1/2}$, and are given by,
\begin{equation}
a_{q\mu}^{\dagger}=(a_{q\mu})^{\dagger}=
i\,n_{q}^{-1/2}\,\sum_{k}c_{k+q\mu}^{\dagger}c_{k\mu}
\end{equation}
It is convenient to introduce spin and charge density operators in place
of $a_{q\uparrow},a_{q\downarrow}$ as follows:
\begin{eqnarray*}
a_{qC}=\frac{1}{\sqrt{2}}\left(a_{q\uparrow}+a_{q\downarrow}\right)\\
a_{qS}=\frac{1}{\sqrt{2}}\left(a_{q\uparrow}-a_{q\downarrow}\right).
\end{eqnarray*}
The corresponding Hermitian fields are,
\begin{eqnarray*}
\varphi_{C}=\frac{1}{\sqrt{2}}\left(\varphi_{\uparrow}+\varphi_{\downarrow}
\right),\\
\varphi_{S}=\frac{1}{\sqrt{2}}\left(\varphi_{\uparrow}-\varphi_{\downarrow}
\right),
\end{eqnarray*}
with commutation relations,
\begin{eqnarray}
\left[\varphi_{C}(x),\,\varphi_{S}(x')\right] &=& 0,\\
\left[\varphi_{C,S}(x),\,\partial_{x'}\varphi_{C,S}(x')\right] 
& = & 2\pi i\delta(x-x').
\end{eqnarray}
In terms of these, we have for the fields and densities,
\begin{eqnarray}
\psi_{C,S}(x) &\equiv& \frac{F_{C,S}}{\sqrt{2\pi a}}e^{-i\varphi_{C,S}(x)},\\
\rho_{C,S}(x) &\equiv& 
                       \psi_{C,S}^{\dagger}(x)\psi_{C,S}(x)\nonumber\\ 
&=& \frac{1}{2\pi}\partial_{x}\varphi_{C,S}(x). 
\end{eqnarray}
\subsection{Bosonized anisotropic Kondo model}
\label{subsec:bosonized-AKM}
We start with the anisotropic Kondo model
\begin{equation}
H_{\rm AKM}=H_{0}+H_{\perp}+H_{\parallel}
\end{equation}
with $H_{0}$ as above and,
\begin{eqnarray*}
H_{\perp} &=&
\frac{J_{\perp}}{2}\sum_{k,k'}\,(c_{k\uparrow}^{\dagger}c_{k'\downarrow}\,S^{-}+c_{k\downarrow}^{\dagger}c_{k'\uparrow}\,S^{+}),\\
H_{\parallel}&=&\frac{J_{\parallel}}{2}\sum_{k,k'}\,(c_{k\uparrow}^{\dagger}c_{k'\uparrow}-c_{k\downarrow}^{\dagger}c_{k'\downarrow})S_{z}.
\end{eqnarray*}
In terms of $\varphi_{C,S}$, we can write 
\begin{eqnarray*}
 H_{0} &=& v_{\rm F}\sum_{q>0}q\,(a_{q\uparrow}^{\dagger}a_{q\uparrow}
+a_{q\downarrow}^{\dagger}a_{q\downarrow})\\
&=& \frac{v_{\rm F}}{2}\int_{-L/2}^{+L/2}\frac{dx}{2\pi}\, :
\left(\partial_{x}\varphi_{C}(x)\right)^{2}+
\left(\partial_{x}\varphi_{S}(x)\right)^{2}:\\
H_{\parallel} & = & \frac{J_{\parallel}}{2} \,S_{z}\,
( \psi_{\uparrow}^{\dagger}(0)\psi_{\uparrow}(0) - 
\psi_{\downarrow}^{\dagger}(0)\psi_{\downarrow}(0) )\\
& = & \frac{J_{\parallel}}{2} S_{z}\, \frac{1}{2\pi}\sqrt{2}\,\partial_{x}\varphi_{S}(0)\\
H_{\perp} &=& \frac{J_{\perp}}{2}\,(\psi_{\uparrow}^{\dagger}(0)
\psi_{\downarrow}(0)\,S^{-}+\psi_{\downarrow}^{\dagger}(0)\psi_{\uparrow}(0)\,S^{\dagger})\\ 
& = & \frac{J_{\perp}}{4\pi a}
\left( e^{i\sqrt{2}\varphi_{s}(0)}F_{\uparrow}F_{\downarrow}^{\dagger}\,S^{-}+
e^{-i\sqrt{2}\varphi_{s}(0)}F_{\downarrow}F_{\uparrow}^{\dagger}\,S^{+}\right)
\end{eqnarray*}

We note that $\varphi_{C}$ (which commutes with $\varphi_{S}$) 
does not couple to the impurity and only gives
a contribution to the kinetic energy. 

\subsection{Equivalence to the Ohmic spin boson model}
\label{subsec:equivalence-ohmic-spin-boson-model}
We show that the unitary transformation 
$U=\exp(i\sqrt{2}S_{z}\varphi_{S}(0))$ applied to $H_{\rm AKM}$ gives 
the spin-boson Hamiltonian, $H_{\rm SBM}$, for Ohmic dissipation, i.e., that
$U H_{\rm AKM} U^{\dagger}=H_{\rm SBM}$. We use the Baker-Hausdorff formula 
$e^{-B}Ae^{B}=A+[A,B]$ with $[A,B]$ a $c$ number and the commutation 
relations for $\varphi_{C},\varphi_{S}$, to obtain,
\begin{eqnarray*}
U H_{0} U^{\dagger} &=&
H_{0}-\sqrt{2}v_{\rm F}S_{z} \left. \frac{\partial\varphi_{S}}{\partial x}
\right| _{x=0}\\
&=& H_{0}-\sqrt{2}v_{\rm F}S_{z}\\
&\times&\sum_{q}\sqrt{\frac{2\pi q}{L}}\,
\left(ia_{qS}+(ia_{qS})^{\dagger}\right)\,e^{-aq/2},\\
U H_{\parallel} U^{\dagger} &=& H_{\parallel} + \mbox{constant},\\
U H_{\perp} U^{\dagger} &=& \frac{J_{\perp}}{4\pi a}\,
(e^{i\sqrt{2}\varphi_{s}(0)}
F_{\uparrow}F_{\downarrow}^{\dagger}\,U S^{-} U^{\dagger}\\
&+& e^{-i\sqrt{2}\varphi_{s}(0)}
F_{\downarrow}F_{\uparrow}^{\dagger}\,U S^{+} U^{\dagger}).
\end{eqnarray*}
On using the identities,
\begin{eqnarray*} 
U S^{-} U^{\dagger} &=& e^{-i\sqrt{2}\varphi_{S}(0)}S^{-},\\
U S^{+} U^{\dagger} &=& e^{i\sqrt{2}\varphi_{S}(0)}S^{+},
\end{eqnarray*}
and the representation, 
$\frac{1}{2}\sigma^{+}=F_{\downarrow}F_{\uparrow}^{\dagger}\,S^{+}$,\, 
$\sigma^{-}=(\sigma^{+})^{\dagger}$ and $\frac{1}{2}\sigma_{z}=S_{z}$,
of the Pauli spin operators, the term $UH_{\perp}U^{\dagger}$ becomes,
\begin{equation}
U H_{\perp} U^{\dagger} = J_{\perp}\,\frac{1}{4\pi
a}\,\sigma_{x}.
\end{equation}
We notice that $a_{qC}$ does not couple to the
impurity so we write $a_{q}=i a_{qS}$ and omit the charge density operators
to obtain the Hamiltonian for the spin density excitations, 
\begin{eqnarray*}
H_{\rm SBM} &=& \frac{J_{\perp}}{4\pi a}\sigma_{x} + 
v_{\rm F}\sum_{q}q\,a_{q}^{\dagger}a_{q}\\
&+&\left(\frac{J_{\parallel}}{4\pi}-v_{\rm F}\right)\sqrt{2}\frac{\sigma_{z}}{2}
\sum_{q>0}\sqrt{\frac{2\pi q}{L}}\,\left(a_{q}+a_{q}^{\dagger}\right)\,e^{-aq/2}.
\end{eqnarray*}
This is precisely the spin-boson model,
\[H_{\rm SBM}=\sum_{q>0}\omega_{q}\,a_{q}^{\dagger}a_{q}-\frac{\Delta_{0}}{2}\sigma_{x}+\frac{q_{0}}{2}\sigma_{z}\sum_{q}\frac{C_{q}}{\sqrt{2m_{q}\omega_{q}}}\,\left(a_{q}+a_{q}^{\dagger}\right)
\]
with $\omega_{q}=v_{\rm F}q$ and a spectral function for the harmonic oscillators,
\[J(\omega)=\frac{\pi}{2}\sum_{q}\frac{C_{q}^{2}}{m_{q}\omega_{q}}\delta(\omega-\omega_{q})
=\frac{2\pi\alpha}{q_{0}^{2}}\,\omega\,
e^{-\frac{\omega}{\omega_{c}}},\]
provided one chooses, 
\begin{eqnarray*}
\frac{C_{q}}{\sqrt{m_{q}}}&=&-\sqrt{\alpha}\frac{2}{q_{0}}\,\left(
\frac{2\pi v_{\rm F}}{L} \right) ^{1/2}\,\omega_{q}\,e^{-\frac{\omega_{q}}{2\omega_{c}}},
\end{eqnarray*}
with a cutoff,
\[\omega_{c}=\frac{v_{\rm F}}{a}.\]
One can identify the parameters,
\[ -\frac{\Delta_{0}}{2}=\frac{J_{\perp}}{4\pi a}\]
and
\[-\sqrt{\alpha}=\frac{J_{\parallel}}{4\pi v_{\rm F}}-1,\]
which, together with the density of states (per spin) of the conduction 
electrons, $\rho=1/2\pi v_{\rm F}$, and the above definition of $\omega_{c}$, 
result in the following identification of the dimensionless couplings of 
the two models,
\begin{eqnarray}
\frac{\Delta_{0}}{\omega_{c}}=-\rho J_{\perp}\\
\alpha=(1-\frac{1}{2}\rho J_{\parallel})^{2}
\label{eq:alpha0}
\end{eqnarray}
The sign of $J_{\perp}$ (or $\Delta_{0}$) plays no role and we may
choose $\Delta_{0}/\omega_{c}=+\rho J_{\perp}>0$.

The precise form of Eq.~(\ref{eq:alpha0}) depends on
the specific regularization scheme used \cite{Guinea1985}. Within the framework of 
Abelian bosonization, the coupling $J_\parallel$ is directly proportional 
to the phase shifts,  $\rho J_\parallel = 4\delta/\pi$ \cite{Emery1992}.
For a finite-band  model with cutoff $\omega_{c}=2D$, the expression
for the phase shift, $\delta$, in terms of $\rho J_{\parallel}$ is
given by $\delta=-\arctan(\pi\rho J_{\parallel}/4)$, and Eq.~(\ref{eq:alpha0})
takes the form:
\begin{equation}
\alpha = (1+\frac{2}{\pi}\delta)^{2}.
\label{eq:alpha} 
\end{equation}
From the above equivalence, one sees that the AKM describes the
physics of the Ohmic spin-boson model for dissipation strengths in the
range $0\leq \alpha \leq 4$ corresponding to $+\infty \leq
J_{\parallel}\leq -\infty$ in the AKM.

\subsection{Equivalence to the spin-fermion model}
\label{subsec:equivalence-spin-fermion-model}
We note that, by replacing the bosonic bath in $H_{\rm SBM}$ above by
a fermionic one, $H_{0}=v_{\rm
  F}\sum_{q>0}qa_{q}^{\dagger}a_q\rightarrow
\sum_{k\mu}\epsilon_{k}c_{k\mu}^{\dagger}c_{k\mu}$,  and making use of the result for the spin density,
\begin{eqnarray*}
S_{z,e}(0)&=&\frac{1}{2}\sum_{k,k'}\,(c_{k\uparrow}^{\dagger}c_{k'\uparrow}-c_{k\downarrow}^{\dagger}c_{k'\downarrow})\\
&=& \frac{1}{2}(c_{\uparrow}^{\dagger}c_{\uparrow}
    -c_{\downarrow}^{\dagger}c_{\downarrow}) = \frac{\sqrt{2}}{4\pi}\partial_{x}\varphi_{S}(0),
\end{eqnarray*}
that we obtain a two-level system coupled to fermions (which we call
the {\em spin-fermion} model),
\begin{eqnarray} 
H_{\rm SFM} &=& -\Delta_{0}S_{x} + H_{0} + J_{z}S_{z}S_{z}^{e}(0),\label{eq:spin-fermion-model}
\end{eqnarray}  
with,
\begin{eqnarray*}
\frac{1}{2}\rho J_{z} &=& 1- \frac{1}{2}\rho J_{\parallel} =
                          \sqrt{\alpha},\\
\frac{\Delta_{0}}{\omega_c} &=& \rho J_{\perp},
\end{eqnarray*} 
or, expressed in terms of the phase shift,
$\delta=\arctan(\frac{\pi\rho J_{z}}{4})$, of conduction electrons
scattering from the potential $J_z/4$:
\begin{eqnarray*}
\alpha &=& (\frac{2\delta}{\pi})^2 =
           (\frac{2}{\pi}\arctan(\frac{\pi\rho J_z}{4}))^2,
\end{eqnarray*}
where $\rho=1/2D$ is the conduction electron density of states per
spin\footnote{One can also start from the spin-fermion model 
(\ref{eq:spin-fermion-model}) and map this directly onto the AKM with the two successive
unitary transformations $U_{1}=\exp[i\sqrt{2}S_{z}\varphi_{S}(0)]$ and
$U_{2}=\exp(i\pi S_{y})$ (A. Rosch, private communication) (see also
Ref.~\onlinecite{Emery1992}).}.
The spin-fermion model describes the physics of the Ohmic two-state
system for dissipation strengths $0\leq \alpha \leq 1$ corresponding
to $0\leq |J_z|\leq +\infty$.
Finally, for a particle-hole symmetric conduction band
$\epsilon_{-k}=-\epsilon_{k}$, making a particle-hole transformation on
the down-spin electrons $c_{k}\rightarrow c_{-k}^{\dagger}$, allows the above
model to be written so that the interaction term couples the
two-level system coordinate $S_{z}$ to the local electronic charge
density $n_{e}(0)=c_{\uparrow}^{\dagger}c_{\uparrow}+c_{\downarrow}^{\dagger}c_{\downarrow}$
instead of the $z$ component of the local electronic spin density. This form of
the model is then suitable for describing the tunneling of atoms in
metallic environments. Further generalizing this model to
include electron-assisted tunneling terms \cite{Vladar1983} leads to
non-Fermi-liquid two-channel Kondo behavior\cite{Zarand2002}.
\begin{table}
\begin{ruledtabular}
\begin{tabular}{ccccc}
\multicolumn{1}{c}{SBM} & 
\multicolumn{1}{c}{AKM} & 
\multicolumn{1}{c}{SFM}& 
\multicolumn{1}{c}{IRLM}\\
\colrule
$\alpha$ & $(1+\frac{2\delta}{\pi})^2,$ & $(\frac{2\delta}{\pi})^2,$&  $\frac{1}{2}(1+\frac{2\delta}{\pi})^2,$\\
  \textit{--}            & $\delta=\tan^{-1}(-\frac{\pi\rho J_{\parallel}}{4})$. &
                                                         $\delta=\tan^{-1}(\frac{\pi\rho J_z}{4})$. &
                                                                   $\delta=\tan^{-1}(-\frac{\pi\rho U}{2})$.\\
$\Delta_{0}$ & $J_{\perp}$ &  $\Delta_{0}$ & $2V$\\
$\omega_{c}$ & $2D$ &  $2D$ & $2D$ \\
$\epsilon$ & $h$ &  $h$ & $\varepsilon_d$\\
\end{tabular}
\end{ruledtabular}
\caption{Parameter correspondence between the Ohmic SBM, and the 
  fermionic anistropic Kondo model (AKM), the spin-fermion model 
  (SFM) and the interacting resonant level model (IRLM). The density 
  of states $\rho=1/2D$. For the spinfull fermionic models, $h$ is a 
  magnetic field which enters the respective Hamiltonians as a term 
  of the form $-hS_{z}=-h\sigma_z/2$. The corresponding term in the 
  IRLM is $\varepsilon_{d}n_d$. 
}
\label{Table3}
\end{table}
\subsection{Equivalence to the IRLM}
\label{subsec:equivalence-IRLM}
We start with the bosonized Kondo model $H_{\rm AKM}$ in terms of the
fields $\varphi_{C,S}$ in Sec.~\ref{subsec:bosonized-AKM} and calculate
$H_{\rm IRLM}=UH_{\rm AKM}U^{\dagger}$ with,
\begin{align}
U & = \exp\left[i\left(\sqrt{2}-1\right)S_{z}\varphi(0)\right].
\end{align}
The calculation is similar to the previous one and we find,
\begin{align}
H_{\rm IRLM} & = v_{\rm F}\sum_{q>0}qa^{\dagger}_{qS}a_{qS}\nonumber\\ 
                 & + \frac{J_{\perp}}{2\pi 
                   a}\left[e^{i\varphi(0)}\sigma^{-}+e^{-i\varphi(0)}\sigma^{+}\right]\nonumber\\
                 & + \left(\frac{J_{\parallel}}{4\pi}\sqrt{2}-(\sqrt{2}-1)v_{\rm 
                   F}\right)\frac{\sigma^{z}}{2}\partial_{x}(0),\nonumber 
\end{align}
with $\sigma^{\pm}$ as defined previously. In terms of the parameters
of the spin-boson model, we have,
\begin{align}
H_{\rm IRLM} & = v_{\rm F}\sum_{q>0}qa^{\dagger}_{qS}a_{qS} + \pi v_{\rm F}\left(1-\sqrt{2\alpha}\right)\sigma^{z}\rho_{S}(0)\nonumber\\
                 & - \frac{\Delta_{0}}{2}\left[e^{i\varphi(0)}\sigma^{-}+e^{-i\varphi(0)}\sigma^{+}\right],\nonumber
\end{align}
where $\rho_{S}\equiv
\rho_{S}(0)=:c^{\dagger}_{S}c_{S:}=\frac{1}{2}(c^{\dagger}_{S}c_{S}-c_{S}c^{\dagger}_{S})+const.$,
is the density of a local spinless fermion field, $\Psi_{S}$:
\begin{align}
\Psi_{S}(0)\equiv c_{S} & = \frac{1}{\sqrt{2\pi a}}F_{S}e^{i\varphi_{S}(0)}.
\end{align}
We drop the index $S$, replace the bosonic bath by a spinless free-fermion Hamiltonian $H_{0}^{\rm F}=v_{\rm
  F}\sum_{k}kc^{\dagger}_{k}c_{k}$, and identify
$c=c_{S}=\frac{1}{\sqrt{L}}\sum_{k}c_{k}$ to obtain the interacting
resonant level Hamiltonian
\begin{align}
  H_{\rm IRLM} & = v_{\rm F}\sum_{k}kc^{\dagger}_{k}c_{k} + V(d^{\dagger}c+c^{\dagger}d)\nonumber\\
               & + \frac{1}{2}\tilde{U}(d^{\dagger}d-dd^\dagger)(c^{\dagger}c-cc^{\dagger}).\nonumber
\end{align}
Here, we have replaced the spin operators, $\sigma^{\pm}$, by fermion
creation and annihilation operators, $d=\frac{1}{2}F_{S}\sigma^{+}$,
$d^{\dagger}=(d)^{\dagger}$, for a localized level at zero energy and
we have made the identification
$\sigma_{z}=d^{\dagger}d-dd^{\dagger}$. It can be seen that the
parameters of the resonant level model are related to those of the
Ohmic spin-boson model by,
\begin{align} 
2\rho \tilde{U} &= (1-\sqrt{2\alpha})\\
-\frac{\Delta_{0}}{2} &= V,  
\end{align} 
where as before $\rho=1/2\pi v_{\rm F}=1/\omega_c$.
In terms of the bare resonant level width $\Gamma = \pi \rho V^{2}$, 
the last equation becomes,
\begin{align}
\pi (\frac{\Delta_{0}}{2\omega_c})^2 & = \frac{\Gamma}{\omega_c}. 
\end{align}
Replacing the potential $\rho \tilde{U}$ by the phase shift $\delta =
-\arctan(\pi \rho \tilde{U})$ gives,
\begin{align}
\alpha & = \frac{1}{2}(1+\frac{2\delta}{\pi})^2,\label{eq:alpha-irlm}
\end{align}
the relation used to connect the IRLM results in this paper to those
for the Ohmic two-state system. Finally, note that in the paper, we
wrote the interaction term in the IRLM as $U (n_d-1/2)(n_0-1/2)$,
implying that $U=2\tilde{U}$ so that $\delta=-\arctan(\pi\rho U/2)$ in
Eq.~(\ref{eq:alpha-irlm}). One sees that the IRLM describes the
physics of the Ohmic two-state system for dissipation strengths $0\leq
\alpha \leq 2$, corresponding to $+\infty \geq U \geq -\infty$ in the IRLM.
Table~\ref{Table3} summarizes the parameter correspondence between the Ohmic spin-boson
model and the three fermionic equivalents discussed in this Appendix.
\section{Additional results}
\label{appendix:additional-results}
We here provide some additional results for thermodynamics
(Sec.~\ref{subsec:additional-results-thermodynamics}) and transient
dynamics (Sec.~\ref{subsec:additional-results-pt}) in support of our
conclusions in the main text.

\begin{figure}[t]
  \includegraphics[width=\columnwidth]{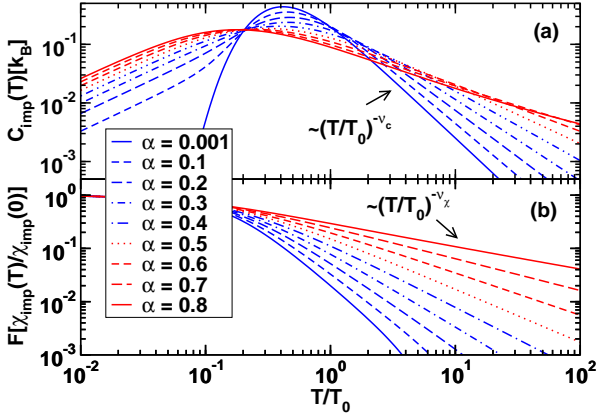}
  \caption 
  {
    Double logarithmic plot for, (a), specific heat $C_{\rm imp}(T)$
    vs $T/T_{0}$, and, (b), the quantity $F[\chi_{\rm 
      imp}(T)/\chi_{\rm imp}(0)]\equiv 4\frac{T}{T_{0}}\frac{\chi_{\rm 
        imp}(T)}{\chi_{\rm imp}(0)}-1$ vs $T/T_{0}$ showing the high-temperature 
    asymptotics $\sim (T/T_{0})^{-\nu_{c,\chi}}$ of these quantities 
    for $T/T_{0}\gg 1$  with $\nu_{c}=\nu_{\chi}=2-2\alpha$. 
  }
  \label{fig:high-T-asymptotics}
\end{figure}
\subsection{Thermodynamics}
\label{subsec:additional-results-thermodynamics}
Figure~\ref{fig:high-T-asymptotics} illustrates the high-temperature
asymptotics of the specific heat and static charge susceptibility of
the IRLM at $T\gg T_{0}$ but still at $T\ll 2D$. The exponents $\nu_{c}=\nu_{\chi}=2-2\alpha$ in
Eqs.(\ref{eq:high-temp-spec}) and (\ref{eq:high-temp-chi}), extracted
by numerical differentiation of the data, are generally well
reproduced for all $\alpha$ (see Table~\ref{Table1}). At sufficiently
high temperature, $T\sim D$, not shown here, non-universal corrections
arise due to the finite bandwidth used in the IRLM. These give rise,
for example, at high enough temperatures to a negative impurity
specific heat at finite $|U|>0$, as found in many other
models\cite{Florens2004,Zitko2009a,Ingold2009,Ingold2012}. 
This does not contradict thermodynamic
stability, since the latter only requires that $C_{\rm tot}(T)\geq 0$
and $C_{0}(T)\geq 0$, which is guaranteed by construction, but the
difference, $C_{\rm imp}(T)$, can take any value. 
\subsection{Transient dynamics}
\label{subsec:additional-results-pt} 
\begin{figure}[t]
  \includegraphics[width=\columnwidth]{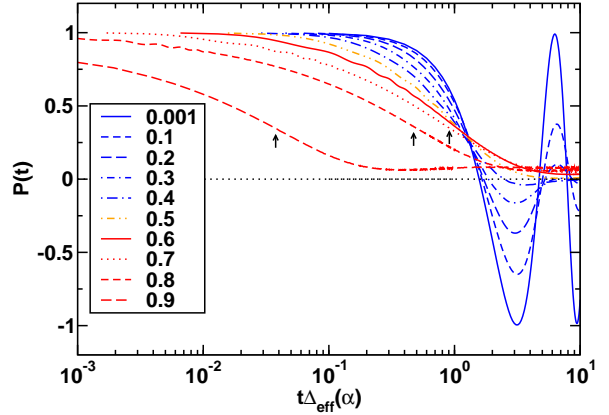}
  \caption{$P(t)$ vs $t\Delta_{\rm  eff}(\alpha)$  for the whole range 
    of $\alpha$ (indicated in the legend). The results are the same as 
    those in Sec.~\ref{subsubsec:comparison-niba} but with additional 
    data extending the time range up to $10/\Delta_{\rm eff}(\alpha)$. 
    The vertical arrows for the strong dissipation cases 
    $\alpha=0.7,0.8$ and $0.9$ indicate the time scale $1/T_0(\alpha)$
    with $T_{0}(\alpha)$ taken from 
    Fig.~\ref{fig:scale-flatband}. The results are obtained via the IRLM with 
    a constant density of states $\rho=1/2D$, $D=1$ and 
    $V/D=\Delta_{0}/\omega_c = 0.1$. 
    NRG parameters: $\Lambda=1.6$, $N_{\rm kept}=2000$, and 
    $n_{z}=32$. 
}
\label{fig:Pt-extra}
\end{figure}
\subsubsection{TDNRG results for $P(t)$ up to $t\sim 10/\Delta_{\rm eff}(\alpha)$}
\label{subsubsec:tdnrg-intermediate-times}
In Fig.~\ref{fig:Pt-extra}, we show results for $P(t)$ extending up to
times of order $t\Delta_{\rm eff}(\alpha)=10$ for all $\alpha$. This
supports our statement in the main text that TDNRG can reach times
of order $1/\Delta_{\rm eff}$ for all $\alpha$. The
scale $\Delta_{\rm eff}(\alpha)$, like the perturbative scale
$\Delta_{r}$, upon which it is based, is incorrect in the Kondo limit
$\alpha\to 1^{-}$ (see Fig.~\ref{fig:scale-flatband},
Table~\ref{Table2}, and Ref.~\onlinecite{Costi1996}). 
In this limit, the dynamics is incoherent and we should expect that
$\Delta_{\rm eff}(\alpha)\approx\gamma_{r}(\alpha)$, where
$\gamma_r(\alpha)$ is the relaxation rate describing the decay of
$P(t)$ at long times. Both scales, $\Delta_{\rm eff}(\alpha)$ and
$\gamma_{r}(\alpha)$, should then be of order the
thermodynamic Kondo scale $T_{0}(\alpha)$ in this limit (see Table~\ref{Table2}). 
However, for $\alpha\gtrsim 0.7$,   it can be seen  from Fig.~\ref{fig:Pt-extra}, 
that $1/\Delta_{\rm eff}(\alpha)$ is not the relevant time scale for 
describing the decay of $P(t)$, being more than an order of magnitude too large for 
$\alpha=0.9$. Instead, $1/T_{0}(\alpha)$, indicated by arrows in
Fig.~\ref{fig:Pt-extra}, sets the time scale for the decay of $P(t)$ for $\alpha\gtrsim 0.7$. 
We see that the TDNRG results appear accurate up to
$t\sim 1/T_{0}(\alpha)$ for $\alpha\gtrsim 0.7$. Since such
time scales are not accessible with the numerically exact TD-DMRG
method for such large $\alpha$, we cannot prove this. One sees,
however, that the long-time limit errors of TDNRG become evident only
at $t\gg 1/T_{0}(\alpha)$ for $\alpha=0.9$. In general, we expect that the TDNRG will
be accurate up to time scales of the order of the inverse decay
rate (relaxation rate), $1/\gamma_{r}(\alpha)$, of the two-level
system, which is the correct time scale for the approach to the
long-time limit. For small $\alpha\ll
1$, this time scale can be much longer than $1/\Delta_{\rm eff}(\alpha)$, as seen
from the NIBA expression for the relaxation rate,
$\gamma_{r}(\alpha)=\sin[\pi\alpha/2(1-\alpha)]\Delta_{\rm
  eff}(\alpha)\approx \pi\alpha\Delta_{\rm eff}(\alpha)/2 \ll
\Delta_{\rm eff}(\alpha)$ for $\alpha\ll 1$. In this limit, TDNRG
simulations remain accurate over many oscillations and can access times
$t\gg 1/\Delta_{\rm eff}(\alpha)$ much longer than the
``intermediate'' time scale $1/\Delta_{\rm eff}(\alpha)$, but still
shorter than the long time scale $1/\gamma_r(\alpha)$. Thus, the
intermediate time scale $1/\Delta_{\rm eff}(\alpha)$ (suitably
corrected for large $\alpha\gtrsim 0.7$) is a
conservative estimate for the longest times up to which TDNRG results
remain free of the errors associated with the approach to the long-time limit.
\begin{figure}[t]
  \includegraphics[width=\columnwidth]{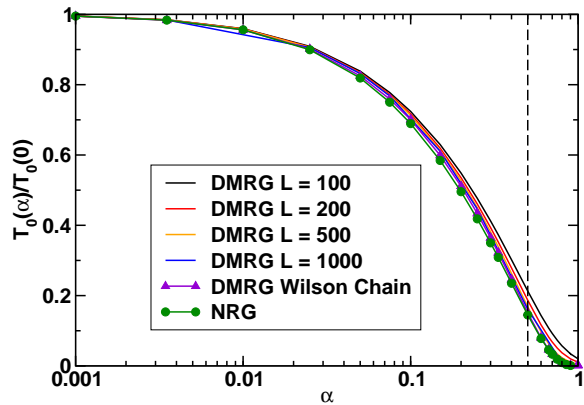}
  \caption 
{Thermodynamic scale $T_{0}(\alpha,L)$ vs $\alpha$ for different 
  tight-binding chain lengths ($L=100,200,500,1000$) in DMRG. Also shown 
  are the DMRG and NRG results using a Wilson chain representation of 
  a semi-elliptic DOS.  The former used $\Lambda=1.6$, while the latter 
  used $\Lambda=4$ and $n_z=8$ (a smaller $\Lambda$ was used in DMRG,
  since $z$-averaging is not used in the latter). In both approaches 
  $T_{0}=1/2\chi(T=0)$ was calculated via a numerical derivative 
$\chi(T=0)=-\partial n_{d}/\partial\veps_d|_{\veps_d=0}\approx -\Delta 
n_d/\Delta\veps_d$ with a sufficiently small increment $\Delta\veps_d$. 
}
\label{fig:t0-semi-elliptic-dmrg-finite-size}
\end{figure}

\subsubsection{Dependence of TD-DMRG results on $L$}
\label{subsubsec:td-dmrg-l-dependence}
\begin{figure}[t]
  \includegraphics[width=\columnwidth]{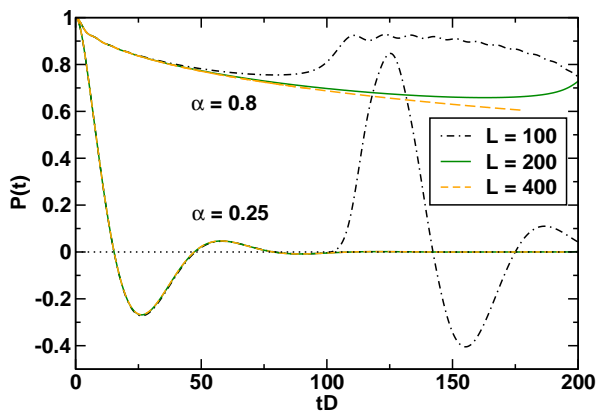}
  \caption 
  {
    Dependence of $P(t)$ on the size $L$ of the tight-binding chain in 
    TD-DMRG for $\alpha=0.25$ and $\alpha=0.8$. For the time scales addressed in the main text $tD\leq 100$ 
    a chain length of $L=200$ suffices for $\alpha\lesssim 0.9$, while 
    $L=100$ would be too small. 
  }
  \label{fig:Ldependence-alpha0p25-alpha0p8}
\end{figure}

Figure~\ref{fig:t0-semi-elliptic-dmrg-finite-size} shows the convergence 
of the low-energy scale $T_0(\alpha, L)$ with increasing 
length $L$ of the tight-binding chain used in the 
DMRG calculations. While the $L = 1000$ results 
show good convergence to the Wilson chain result, this length of chain would be numerically very 
expensive within a TD-DMRG calculation. However, due to the 
“Lieb-Robinson bound” \cite{Lieb1972} 
one can employ shorter chains ($L=200$) for the transient dynamics 
for the time scales shown as explained in the main text ($tD\lesssim 100$). 
This is illustrated in Fig.~\ref{fig:Ldependence-alpha0p25-alpha0p8}, which 
shows results for $P(t)$ at $\alpha=0.25$ and $0.8$. Converged results out 
to times $tD=100$ are achieved for $L\geq 200$. In contrast, the chain 
of length $L=100$ exhibits exhibits finite-size effects due to reflections from the end 
of the chain reaching the impurity on time scales $tD\sim
100$. Similar effects arise in other approaches, such as in studies of
the transient dynamics of the Anderson model within an optimal basis
approach, when the number of orbitals for the reservoir is not large
enough \cite{Lin2015}. By explicitly checking the $L$ dependence of
the TD-DMRG results, we confirmed that the results in the text 
for all $\alpha$ were converged for the times shown for $L=200$.
\subsubsection{Ringing effect}
\label{subsubsec:ringing-effect}
\begin{figure}[t]
  \includegraphics[width=\columnwidth]{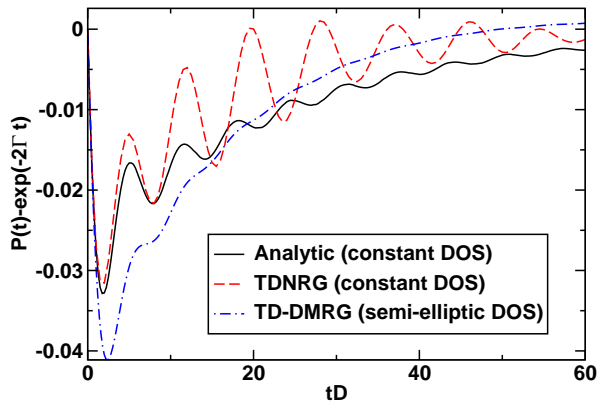}
  \caption 
  {
    $P(t)-\exp(-2\Gamma t)$ at $\alpha=1/2$ with $\Gamma=0.02D$ and 
    $P(t)$ calculated, (i), analytically 
    for a constant DOS $\rho(\omega)=1/2D, -D\leq 
    \omega\leq+D$ (black solid line), (ii), using TDNRG with a 
    constant DOS (red dashed curve), and, (iii), within TD-DMRG for a 
    semi-elliptic DOS using a tight-binding chain of 
    length $L=200$ (blue dashed-dotted line).  
  }
  \label{fig:wiggles}
\end{figure}

Figure~\ref{fig:wiggles} illustrates the ringing effect mentioned in the
text for the exactly solvable case of $\alpha=1/2$.
The analytic result for $P(t)$ can be evaluated
for both a finite cutoff $\omega_c=2D$ and in the so-called scaling
limit $\omega_c\to \infty$, 
with the low-energy scale $\Delta_{\rm  
  eff}(\alpha=1/2)=2\Gamma$ being kept constant\cite{Anders2006}. The scaling limit
result at $\alpha=1/2$, which is also the NIBA result, is simply $P(t)=\exp(-2\Gamma
t)$. A finite high-energy cutoff $\omega_c=2D$ introduces a new 
ultrashort time scale $1/D$ in the response of the system to a sudden
switching and modifies the scaling limit result at short times. 
In Fig.~\ref{fig:wiggles}, we show the corrections to the scaling
limit result, i.e., $P(t)-\exp(-2\Gamma t)$. This measures
the magnitude and decay of the ringing oscillations. As expected,
their frequency is set by $D$. They decay as $1/t^3$ at long times
$t\gg 1/D$. The TDNRG results exhibit similar oscillations, as do
the TD-DMRG results. The latter are more damped, presumably
because the use of a semi-elliptic DOS, corresponding to a
tight-binding chain, is a more accurate description of the continuum
bath than the Wilson chain of the TDNRG. Similary, the FRG data shown in
Fig.~\ref{fig:compare-FRGNRG-semi-elliptic} exhibit a much weaker
ringing effect, than the corresponding TDNRG data, again, presumably
due to the use of a continuum bath in the FRG approach.
\begin{figure}[t]
  \includegraphics[width=\columnwidth]{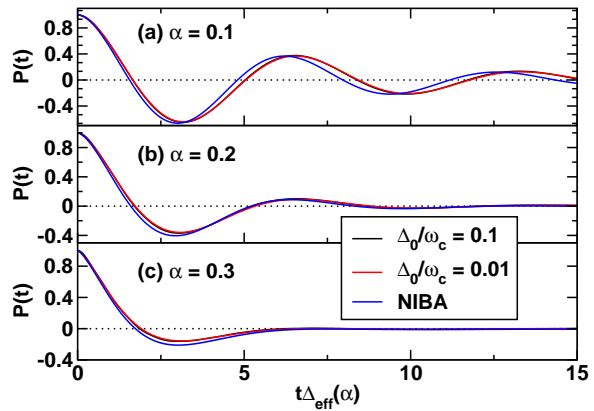}
  \caption 
  {Dependence of TDNRG results for $P(t)$ vs $t\Delta_{\rm eff}(\alpha)$
    on the ratio 
    $V/D=\Delta_{0}/\omega_{c}$ for, (a), $\alpha=0.1$, 
    (b), $\alpha=0.2$, and, (c), $\alpha=0.3$.  
    A constant density of states was used for the IRLM with $V/D=\Delta_{0}/\omega_{c}=0.1$,
    and $V/D=\Delta_{0}/\omega_{c}=V/D=0.01$. The NIBA results, which  
    are independent of $\Delta_0/\omega_{c}$, are also shown. 
    NRG parameters as in Fig.~\ref{fig:niba-alpha0p001alpha0p01}. 
  }
  \label{fig:v-dependence}
\end{figure}
\subsubsection{Dependence of TDNRG results on
  $\Delta_{0}/\omega_{c}=V/D$, $N_{\rm kept}$, and $\Lambda$}
\label{subsubsec:tdnrg-dependences}
 The NIBA is usually believed to provide a reasonable description of
the short to intermediate time dynamics of $P(t)$ for 
$0\lesssim \alpha \lesssim1/2$ \cite{Leggett1987,Weiss2008}.
In Sec.~\ref{subsubsec:comparison-niba} we found significant deviations
between the NIBA and TDNRG results, particularly for $0.1\lesssim
\alpha\lesssim 0.4$. In order to exclude the possibility that this difference
is due to non-converged TDNRG results (with respect to the number of
retained states, $N_{\rm kept}$, or the value of $\Lambda$) or that
the TDNRG results are not sufficiently deep in the scaling limit,
$\Delta_{0}/\omega_c\ll 1$,  we
performed various additional checks, as we now describe. In Fig.~\ref{fig:v-dependence}, we show the
effect of decreasing $\Delta_0/\omega_c=0.1$ by a factor of $10$
on the TDNRG results for $P(t)$ at $\alpha=0.1,0.2$, and $0.3$. We see
that, on the relevant time scales $t \sim 1/\Delta_{\rm
  eff}(\alpha)$, the change in the TDNRG results is insignificant and
much smaller than the observed differences with the NIBA results. This
suggests that these differences are not due to the finite 
$\Delta_0/\omega_c$ used in the TDNRG calculations, but represent
quantitative differences between the TDNRG and NIBA results. We note
that local quantities such as $n_d(t)$ in the IRLM [and hence $P(t)=2n_{d}(t)-1$]
are less sensitive to band-edge effects than extensive quantities, such
as specific heats $C_{\rm tot}(T)$ and $C_{0}(T)$, and differences
thereof such as $C_{\rm imp}(T)=C_{\rm tot}(T)-C_{0}(T)$. Thus, while in Sec.~\ref{subsec:thermodynamics}
universal results for $C_{\rm  imp}(T)$ required using
$\Delta_0/\omega_c=V/D\ll 1$, for the local quantity $P(t)$, values of
$\Delta_{0}/\omega_c=V/D=0.1$ suffice to obtain universal results 
on the time scales of interest. Non-universal effects in $P(t)$, due to
the finite cutoff $D$ in the IRLM, only appear on time scales
$t\lesssim 1/D$.

\begin{figure}[t]
\includegraphics[width=\columnwidth]{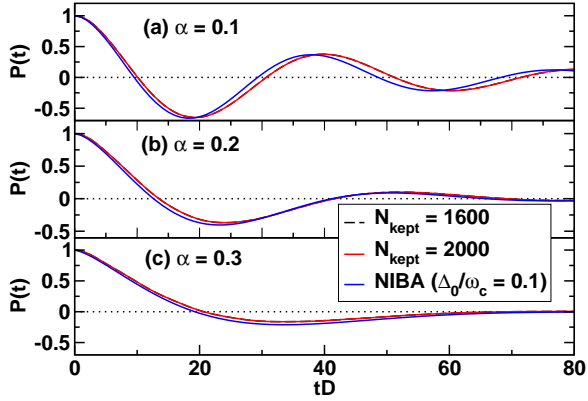}
  \caption 
{Dependence of the TDNRG result for $P(t)$ on the number of kept 
  states $N_{\rm kept}$ for, (a), $\alpha=0.1$, (b), $\alpha=0.2$, and,
  (c), $\alpha=0.3$. A constant density of states was used for the  
  IRLM with  $V/D=\Delta_{0}/\omega_{c}=0.1$. 
  Also shown is the NIBA result.  
  NRG parameters as in Fig.~\ref{fig:niba-alpha0p001alpha0p01}. 
}
\label{fig:state-dependence-flatband}
\end{figure}

The dependence of the TDNRG results on the number of states
is also seen to be insignificant for $N_{\rm kept}\gtrsim 1600$ (see 
Fig.~\ref{fig:state-dependence-flatband}). A strong dependence
of the TDNRG results on the number of retained states is only found if this
number is taken to be too small, e.g., of the order $100$ instead of
the order $1000$ as used in our calculations. 
\begin{figure}[t]
  \includegraphics[width=\columnwidth]{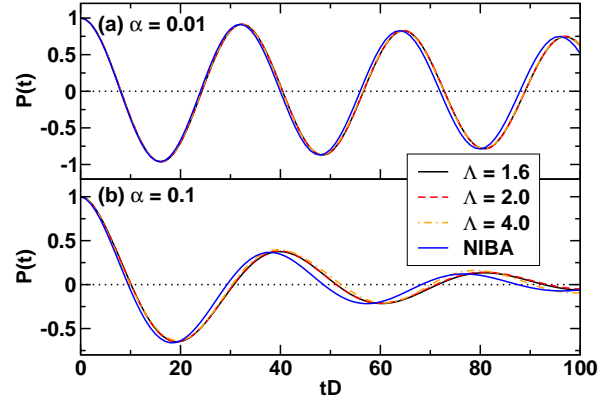}
  \caption 
  {Dependence of the TDNRG results for $P(t)$ vs $t$ on the logarithmic 
    discretization parameter $\Lambda$ using $N_{\rm kept}=2000$ and 
    $n_{z}=32$, for, (a) $\alpha=0.01$, and, (b), $\alpha=0.1$. 
    A constant density of states was used for the IRLM with 
    $V/D=\Delta_{0}/\omega_{c}=0.1$.    Also shown is the NIBA 
    result. 
  }
  \label{fig:lambda-dependence}
\end{figure}

Finally, Fig.~\ref{fig:lambda-dependence} shows the dependence of the TDNRG
results on decreasing the logarithmic discretization parameter
$\Lambda$ for $\alpha=0.01$ and $0.1$. While the use of a
smaller $\Lambda$,  better approximates the continuum bath at
$\Lambda\to 1$, and hence can improve the long-time limit
of TDNRG calculations \cite{Anders2006,Nghiem2014a,Nghiem2014b}, in
practice one cannot take the limit $\Lambda\to 1$, since the loss in
accuracy coming from the NRG truncation eventually outweighs any
gains in approaching the continuum limit at $\Lambda\to 1$. The value
$\Lambda=1.6$, used here, therefore represents a compromise between
these two effects. One sees from  Fig.~\ref{fig:lambda-dependence} that
the results for $\Lambda=2$ are almost indistinguishable to the
$\Lambda=1.6$ results, whereas those for $\Lambda=4$, which corresponds to a coarser description of the bath,
start to show some deviation from the smaller $\Lambda$ results. These
deviations are particularly noticeable for the $\alpha=0.1$ case. This
motivated our use of $\Lambda=1.6$ for all TDNRG calculations in this paper.

\bibliography{lit}
\end{document}